\documentclass[lettersize,journal]{IEEEtran}
\usepackage{amsmath}
\usepackage{subcaption}
\usepackage{graphicx}
\usepackage{algorithm}
\usepackage{algcompatible}
\usepackage{algorithmicx}
\usepackage{algpseudocode}
\usepackage{color}
\usepackage{url}
\usepackage{booktabs}
\newcommand{\sect}{Section~} 
\newcommand{\fig}{Fig.~} 
\newcommand{\systemname}{VibOmni}
\usepackage{balance}  
\newcommand{\blue }{\color{black}}

\newcommand{\new}[1]{{\color{black}#1}}

\usepackage[colorinlistoftodos]{todonotes}
\begin{document}

\title{\systemname{}: Towards Scalable Bone-conduction Speech Enhancement on Earables}

\author{
  Lixing He, Yunqi Guo,~\IEEEmembership{Member,~IEEE}, Haozheng Hou, Zhenyu~Yan,~\IEEEmembership{Member,~IEEE}
  % , Guoliang~Xing,~\IEEEmembership{Fellow,~IEEE}
  % Jane~B.~Doe, 
  % and~Robert~C.~Lee,~\IEEEmembership{Member,~IEEE}% <- No space before this line
  \thanks{L. He, Y. Guo, H. Hou, and Z. Yan are with the department of information engineering, The Chinese University of Hong Kong, Hong Kong.}
  \thanks{\new{This paper is an extended version of our prior work appeared in the proceeding of ACM MobiSys 2023~\cite{he2023towards}.}}
}


\markboth{IEEE TRANSACTIONS ON MOBILE COMPUTING,~Vol.~14, No.~8, August~2021}%
{Shell \MakeLowercase{\textit{et al.}}: A Sample Article Using IEEEtran.cls for IEEE Journals}


\maketitle

\begin{abstract}
Earables, such as True Wireless Stereo earphones and VR/AR headsets, are increasingly popular, yet their compact design poses challenges for robust voice-related applications like telecommunication and voice assistant interactions in noisy environments. Existing speech enhancement systems, reliant solely on omnidirectional microphones, struggle with ambient noise like competing speakers. To address these issues, we propose \systemname{}, a lightweight, end-to-end multi-modal speech enhancement system for earables that leverages bone-conducted vibrations captured by widely available Inertial Measurement Units (IMUs). \systemname{} integrates a two-branch encoder-decoder deep neural network to fuse audio and vibration features. To overcome the scarcity of paired audio-vibration datasets, we introduce a novel data augmentation technique that models Bone Conduction Functions (BCFs) from limited recordings, enabling synthetic vibration data generation with only 4.5\% spectrogram similarity error. Additionally, a multi-modal SNR estimator facilitates continual learning and adaptive inference, optimizing performance in dynamic, noisy settings without on-device back-propagation. Evaluated on real-world datasets from 32 volunteers with different devices, \systemname{} achieves up to 21\% improvement in Perceptual Evaluation of Speech Quality (PESQ), 26\% in Signal-to-Noise Ratio (SNR) and about 40\% WER reduction with much less latency on mobile devices. A user study with 35 participants showed 87\% preferred \systemname{} over baselines, demonstrating its effectiveness for depolyment in diverse acoustic environments.
\end{abstract}


\section{Introduction}
\label{sec:intro}

Earables are smart devices designed to be worn on users' heads or ears, including products such as True Wireless Stereo (TWS) earphones, VR/AR headsets, and smart glasses. These devices are equipped with various sensors and support a range of applications, including virtual and augmented reality (VR/AR), motion recognition, and voice assistants. TWS earphones, like the Apple AirPods series, have significantly contributed to the growth of the earables market, which has become the largest category among all wearable devices, with an estimated shipment of over 273 million units worldwide in 2023 \cite{wearable-stat}.
Manufacturers are continuously enhancing earables by adding new functionalities. For instance, many earphones and headphones now feature active noise cancellation (ANC) to improve the listening experience. Additionally, voice-related applications that utilize the microphones in earables are among the most commonly used features.

One of the key applications of earables is making phone calls, which an increasing number of people are utilizing. Users can also interact with voice assistants, such as Siri and Alexa, through voice commands. However, the speech quality on earables often falls short due to several challenges:
First, most earables use omnidirectional microphones that capture sound from all directions, which can result in unwanted environmental noise.
Second, while many earables are equipped with multiple microphones to create an array for noise reduction via beamforming, the proximity of the microphones to one another can hinder their performance in effectively isolating the user's voice.
Third, the speech audio is significantly attenuated by the time it reaches the earables, as they are typically positioned far away from the user's mouth.

% Existing approach
Various approaches have been developed for speech enhancement. Signal processing-based methods \cite{ephraim1995signal} remove noises based on their statistical models. However, these approaches cannot handle complex environments. Microphone beamforming \cite{yousefian2014hybrid,ji2017coherence} removes noises based on their directions. But they fail to distinguish the speaker's voice and noise when the microphones are placed too close. Several research \cite{subakan2021attention, zhang2021microphone, chatterjee2022clearbuds} adopt deep neural networks (DNNs) to improve speech quality. However, their performance varies when the domain changes.  
Except for the audio-only solution, several works leverage other modalities like contact sensors \cite{gupta2018precision}, vibration sensor \cite{maruri2018v}, camera \cite{ gao2021visualvoice}, mm-Wave Radar \cite{ozturk2021radiomic, li2020vocalprint, liu2021wavoice}, ultrasonic sensing \cite{sun2021ultrase, zhang2021sensing}, and Lidar \cite{sami2020spying}, which introduce additional hardware requirements or user overhead to existing earables.

% introduce vibration
Motivated by the above works, we envision finding a modality that: 1) can be obtained with existing earables without significant modification. 2) has a close correlation with the user's speech and may not be influenced by noises. Since earables are well connected to the user's head, it becomes a desired position to obtain the vibration transmitted to the user's head. The vibration, which is also known as bone conduction vibration, is less influenced by the ambient sound and mainly depends on the clear speech from the user's vocal tract.
Fortunately, the existing sensors of commercial earables include an Inertial Measurement Unit (IMU), whose original function is to track the head pose. At the same time, the IMU (including accelerometer and gyroscope) can also capture the subtle vibration on the head, which corresponds to the bone conduction vibration. Considering the mainstream sampling rate of IMU is about $1.6~\text{kHz}$, the bandwidth ($800\text{Hz}$) overlaps with the lower part of the frequency range of human speech. 

\begin{figure}
    \centering
    \includegraphics[width=0.75\columnwidth]{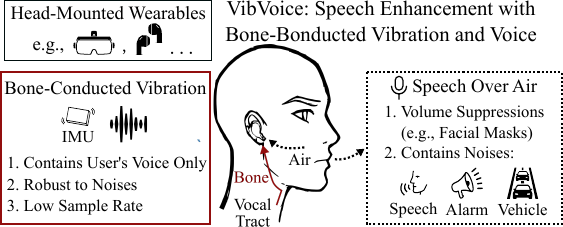}
    \caption{\systemname{} enhances speech quality of head-mounted wearables by extracting the user's clear voice from the bone-conducted vibrations.}
    \label{fig:intro}
    \vspace{-1em}
\end{figure}

To enhance speech quality in earables, we propose \systemname{}, a low-latency, end-to-end multi-modal speech enhancement system using bone-conducted vibrations and audio, as shown in Fig. \ref{fig:intro}. Optimized for mobile devices, \systemname{} leverages vibration's noise-robustness and audio's rich information, addressing three key challenges:
\begin{itemize}
    \item \textbf{Multi-Modal Data Fusion}: Microphone audio, with its higher sample rate, provides richer information than vibration data, risking DNN overfitting and neglect of vibration input. A specialized training strategy with dual loss functions is needed to balance both modalities.
    \item \textbf{Scarcity of Paired Data}: Collecting paired audio-vibration datasets for head-mounted wearables is labor-intensive. Existing public datasets lack paired data, and large-scale collection involves significant overhead from diverse volunteers and extensive annotated recordings.
    \item \textbf{Real-World Deployment}: Dynamic interferences complicate deploying deep learning speech enhancement. Limited and variable-quality user-collected data may hinder training a robust model, even with rapid data collection.
\end{itemize}
% \begin{itemize}
%     \item 
%     \noindent\textbf{Fusion of data with different modalities and sample rates.}
%     Microphone audio has much richer information than vibration data due to the higher sample rate. As a result, the DNN is prone to overfitting and may ignore the data input of vibration. 
%     Therefore, during the training of DNN, we require a special training strategy to avoid overfitting and two dedicated losses to balance the weights of the two modalities.
    
%     \item 
%     \noindent\textbf{Lack of paired labeled vibration and audio.}
%     Collecting a paired multi-modal dataset is labor-intensive. Although there are public datasets of either vibration or audio, none have the paired data on head-mounted wearables. 
%     Collecting such a paired dataset on a large scale can introduce excessive overheads, such as recruiting diverse volunteers and recording hundreds of hours of audio with annotations. 

%     \item 
%     \noindent\textbf{Practical concern for deployment.} 
%     Deploying deep learning speech enhancement in the real world is non-trivial due to the dynamic interferences.
%     Even though some users may tolerate a rapid data collection phase before use, the amount and quality of the collected data are not guaranteed for training a high-performance model.

% \end{itemize}

First, unlike previous work like audio-only speech enhancement or target speaker extraction that relies on time-invariant speaker embeddings, our multi-modal speech enhancement network (\S\ref{sec:network}) employs two encoders to extract and concatenate features from both modalities, with a projection layer aligning frequency dimensions due to differing sampling rates. A DPRNN module separates speech at the feature level and utilizes two decoders to estimate the full-band and low-band speech (to avoid the collapse of multi-modal learning), which is trained by SISNR loss.
The whole model is constructed by causal convolutions and unidirectional RNNs for frame-by-frame inference. 

Second, to overcome the scarcity of paired audio-vibration datasets, \systemname{} employs an innovative pre-training strategy using Bone Conduction Functions (BCFs) (\S\ref{sec:bcf}). By estimating BCFs from limited recordings and applying them to public datasets like LibriSpeech \cite{panayotov2015librispeech}, we generate synthetic vibration data, achieving a low 4.5\% error in spectrogram similarity. This approach, enhanced by cubic interpolation and Gaussian modeling, ensures diverse training data and improves model generalization across users and conditions, reducing the need for extensive user-specific fine-tuning. These advancements enable \systemname{} to bridge the gap between training and real-world testing, particularly in noisy, dynamic environments.

\new{
Lastly, to tackle real-world speech enhancement challenges, \systemname{} integrates advanced continual learning and adaptive inference mechanisms, as detailed in Sections \ref{sec:continual_learning} and \ref{sec:adaptive_inference}. We propose a multi-modal SNR estimator to distinguish clean audio from noise, enabling a continual self-supervised learning approach that leverages only noisy, in-the-wild mixtures. Additionally, an adaptive inference mechanism dynamically adjusts the model's depth based on estimated noise levels, optimizing computational efficiency. These strategies ensure effective, privacy-preserving adaptation without requiring on-device back-propagation, making \systemname{} ideal for resource-constrained mobile devices in diverse acoustic environments.

We evaluate \systemname{} on both add-ons of earables and the development board, with paired vibration and audio from 32 volunteers in total. We evaluate \systemname{}'s performance with both a synthetic noise dataset and in-the-wild noise. Specifically, \systemname{} achieves the best performance with 31 times less latency on mobile devices than the two strong baselines, which is promising to be deployed on mobile devices. In addition, data augmentation can reduce the requirement of paired data by more than 72 times.
Besides, our proposed SNR estimator obtains 3dB errors in average, which is much better than the audio-only baseline. With the estimated SNR, the adaptive training and testing achieve up to 3dB boost of SNR improvement and prune the unnecessary computation.
For the perceptual evaluation, we recruit 35 volunteers for a user study in which \systemname{} is preferred by 87$\%$ users compared to the baseline.
}

We summarize our contributions as follows:
\begin{itemize}
    \item We develop a multi-modal deep neural network that extracts clean speech from noisy audio with assistance from the vibration signal, and employs a lightweight architecture suitable for low-latency execution on mobile devices.
    \item We introduce a novel data augmentation approach for modeling the Bone Conduction Function, enabling the augmentation of paired vibration and audio data by leveraging a large public dataset. 
    \item We propose a multi-modal SNR estimator to enhance adaptation in training and testing phases, enabling continual learning and adaptive inference, which boosts the effectiveness and efficiency.
\end{itemize}

\section{Related work}\label{sec:relatedwork}

\subsection{Speech Enhancement}

\paragraph{Audio-only enhancement}
Traditional speech enhancement assumes signal stationarity, speech-noise independence in the time-frequency domain \cite{ephraim1995signal}, or uses microphone arrays for beamforming to enhance audio quality by leveraging arrival time differences \cite{yousefian2014hybrid,ji2017coherence}. These methods struggle with dynamic noises without prior knowledge.
Recent DNN-based approaches \cite{subakan2021attention, zhang2021microphone, chatterjee2022clearbuds} enhance speech by capturing voice and noise features from large datasets, supporting either single-channel \cite{subakan2021attention} or multi-channel audio with \cite{zhang2021microphone}. Importantly, ClearBuds \cite{chatterjee2022clearbuds} uses DNNs for stereo audio from Earbuds but needs dedicated devices. 

\paragraph{Multi-modal enhancement}
Except for using audio only, any other modalities that are correlated to the speech can be leveraged for speech enhancement.
Previous work \cite{gao2021visualvoice} uses camera videos to correlate audio-visual data for speech enhancement and separation via deep learning and cross-modal embeddings. However, cameras are not always available in daily scenarios.
Differently, works like \cite{sun2021ultrase, zhang2021sensing} use smartphones to emit inaudible acoustic signals ($>$17 kHz) and capture lip-reflected echoes to enhance noisy audio. Others works \cite{liu2021wavoice} combine mmWave radar with microphones for speech recognition.
Recent studies use bone conduction sensors or accelerometers with microphones for multi-modal speech enhancement \cite{wang2022fusing, wang2022multi, tagliasacchi2020seanet}. However, their scalability is questionable without an explicit understanding of bone-conduction.

\paragraph{Target speech enhancement}
Different from introducing a new modality, a predefined target can be applied to speech enhancement as well. Compared to incorporating a new modality (e.g., vision, vibration) as the target, the above target can be obtained in a one-time effort.
Previous work considers features that correlate well with the target sound as the condition, such as the speaker embeddings \cite{veluri2024look}, sound class \cite{veluri2023semantic}, or proximity to the user \cite{chen2024hearable}. 
However, all of them need manual involvement of the user, such as setting a key parameter (the distance to define proximity).

\subsection{Acoustic Sensing on Earables}

\paragraph{In-ear sensing}
In-ear sensing in ANC earphones leverages the built-in microphone and speaker to detect vital signs and enable various applications. The occlusion effect, where the ear canal amplifies low-frequency sounds, supports step counting, activity recognition, and gesture recognition, as demonstrated in studies like Oesense \cite{ma2021oesense}. Additionally, active sensing (using both the microphone and the speaker) using ultrasonic waves emitted by the speaker and analyzing their reflections enables applications such as authentication \cite{wang2021eardynamic}, silent speech interfaces \cite{dong2024rehearsse}.

\paragraph{Out-ear sensing}
Sensing the details of the face by earables primarily relies on active sensing due to the absence of the occlusion effect present in in-ear sensing. Research focuses on analyzing facial expressions \cite{song2022facelistener}, enhancing speech \cite{duan2024earse}, authentication \cite{duan2024f2key}. Smart glasses, with their distinct microphone and speaker placement, support applications like authentication \cite{li2024sonicid}, facial expression analysis \cite{li2024eyeecho}, and pose estimation \cite{mahmud2023posesonic}. 

% Additionally, bone-conduction sound, which captures unique head-related features, enables authentication \cite{shin2024skullid}, vital sign monitoring \cite{he2024hcr}, and gesture recognition \cite{yang2024maf}.

\section{Background and Motivation Study}
\label{sec:background}

\subsection{Bone Conduction Vibration}
In this paper, we focus on the vibration transmission from the vocal cord to the skull and refer to it as bone conduction vibration as follows: 
\begin{equation}
  s_{vib} = f(s_{speech}) + \epsilon_{vib} 
  \quad
  s_{mic} = s_{speech} + \epsilon_{mic}
  \label{math:transfer_function}
\end{equation}
where $s_{vib}$ and $s_{mic}$ are the raw data captured by the accelerometer and the microphone, respectively; $s_{speech}$ denotes the ground-truth (clean) speech audio; $\epsilon_{vib}$ and $\epsilon_{mic}$ are environmental noises captured by the accelerometer and the microphone, respectively; and $f$ is the Bone Conduction Function (BCF).
The noiseless feature of BCF has been discovered \cite{tagliasacchi2020seanet}, which has also been applied in extremely harsh environments like battlefields or underwater \cite{BoneCond30:online}. However, there are still challenges to leveraging it for speech enhancement on earables: 1) the propagation through the head is complicated, resulting in an unstable response even for the same location \cite{chang2016development, won2005estimating}. 2) They are too expensive (e.g., 60 US dollars) and not available on commercial earables.

\subsection{Vibration Sensing Platform}
\label{sec:EarSense}
\begin{figure}
  \begin{subfigure}{.64\columnwidth}
  \centering
    \includegraphics[width=0.75\textwidth]{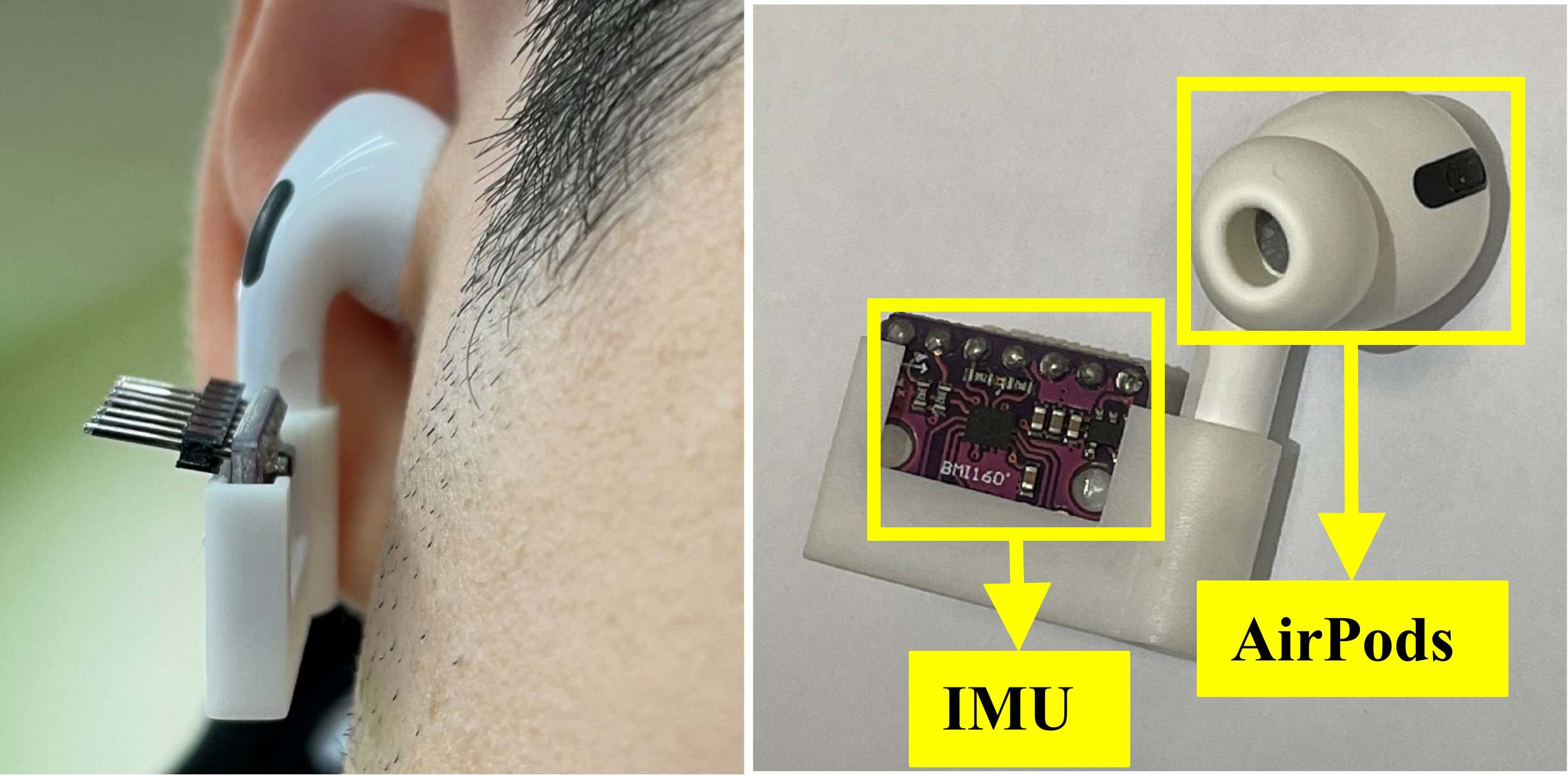}
    \caption{\emph{EarSense} with an Airpods Pro.}
    \label{fig:earsense-head}
  \end{subfigure}
  \begin{subfigure}{0.35\columnwidth}
    \centering
    \includegraphics[width=0.75\textwidth]{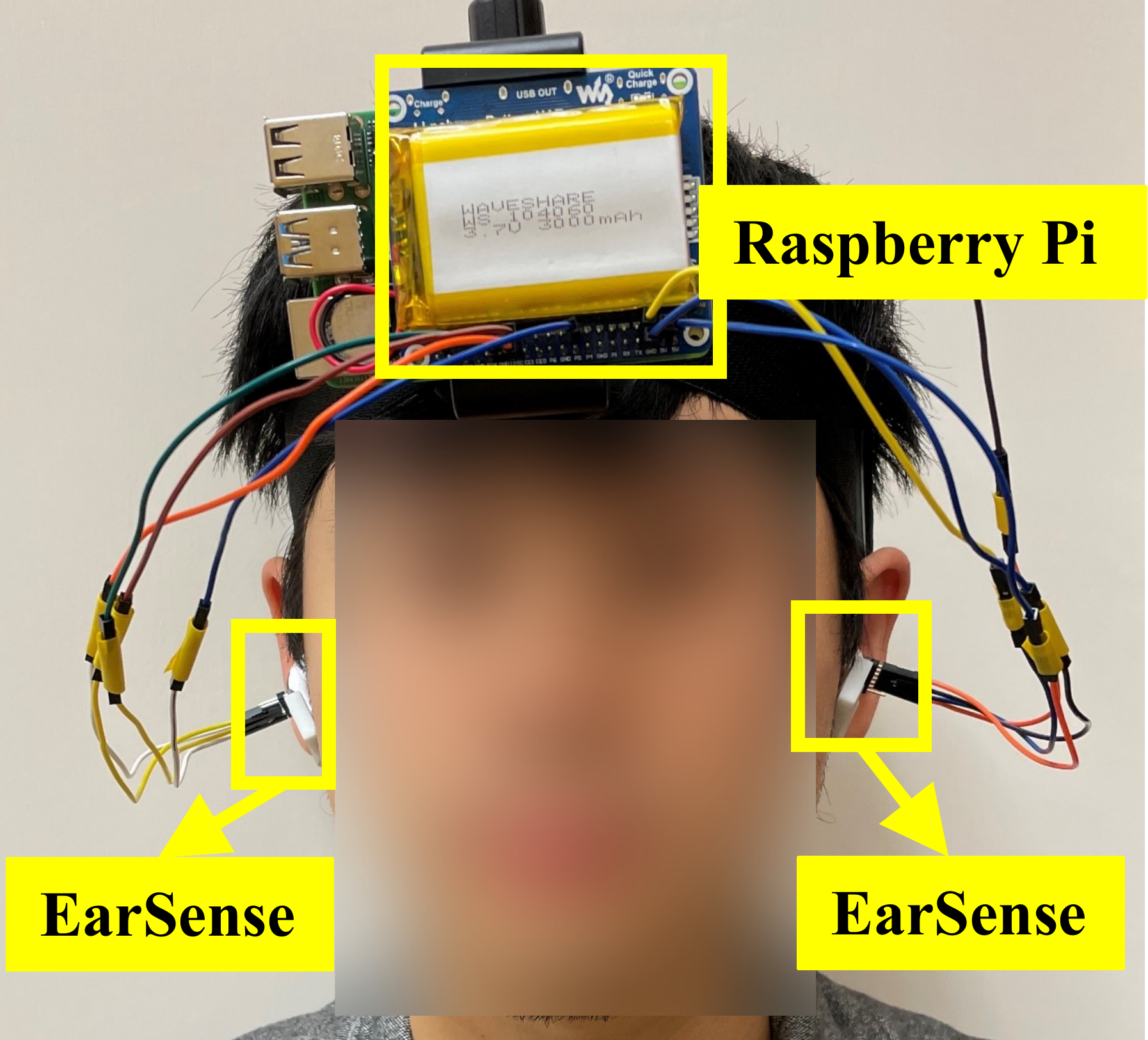}
    \caption{Test Setup.}
    \label{fig:earsense-people}
  \end{subfigure}
  \caption{\emph{EarSense} is an open-sourced data collector attachable to commercial earabless for vibration sensing.}
    \vspace{-1em}
  \label{fig:earsense}
\end{figure}

Most commercial earables do not provide APIs to collect raw acceleration data. Recently, Apple provided APIs \cite{CMHeadph3:online} to collect motion data from AirPods earphones at $100\,\text{Hz}$, which is too low for speech recording and doesn't fully utilize commercial IMU.
Hence, it is desirable to develop a new sensing platform \textit{EarSense} that can collect the acceleration and acoustic data synchronously on commercial earables.
Fig. \ref{fig:earsense} presents the prototype, which is a sensing platform with a 3D-printed enclosure and a Bosch BMI-160 IMU sensor \cite{Inertial63:online}. 
We can attach our platform on commercial earables like AirPods Pro to capture bone-conducted vibration, as shown in Fig. \ref{fig:earsense-head}. Specifically, two EarSense units connect to a battery-powered Raspberry Pi (RPi) for data collection, and the accelerometer data is streamed by I2C.
The microphone and accelerometer sample at 16 kHz and 1.6 kHz, respectively.
Specifically, we only access mono audio recording due to AirPods Pro limitations.

\subsection{Measurement Study}
\label{sec:measurement}

\begin{figure}
    \begin{subfigure}{0.48\columnwidth}
        \centering
        \includegraphics[width=1\textwidth]{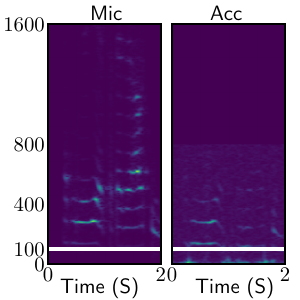}
        \caption{The user is talking with no noise as the reference recording.}
        \label{fig:nospeaker}
        \end{subfigure}
        \hspace{.5em}
    \begin{subfigure}{0.48\columnwidth}
        \centering
        \includegraphics[width=1\textwidth]{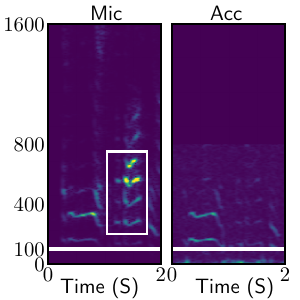}
        \caption{The user is talking with a competing speaker.}
        \label{fig:speaker}
         \end{subfigure}
      \caption{Microphone and accelerometer recording.}
      \vspace{-1em}
  \label{fig:competing_speaker}
  
\end{figure}

\begin{figure}
   \begin{minipage}[t]{0.45\columnwidth}
       \centering
        \includegraphics[width=1\linewidth]{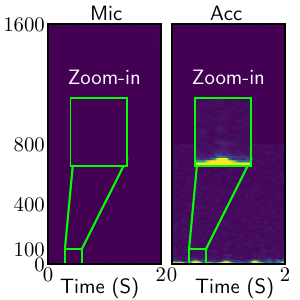}
    \caption{The user is walking with no noise as the motion reference.}
        \vspace{-1em}

        \label{fig:walk}
    \end{minipage}
    \hfill
\begin{minipage}[t]{0.45\columnwidth}
      \centering
    \includegraphics[width=1\linewidth]{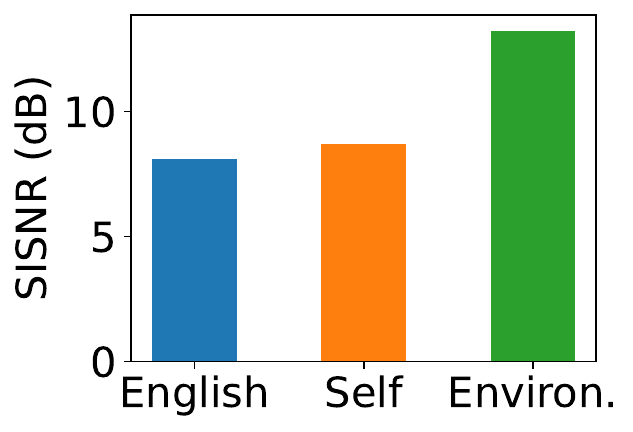}
    \caption{Enhancement performance on different kinds of noises.}
        \vspace{-1em}
    \label{fig:cross_noise}
\end{minipage}
\end{figure}

We conducted an experiment where the user (with \textit{EarSense}) was in a meeting room ($10m^2$), speaking when we recorded both the audio and vibration. Note that we apply L2-Norm to the three axes of acceleration to extract vibration intensity while reducing effects from wearing position and motion.

\noindent\textbf{Competing speaker} refers to the scenario when there is another speaker around the user. We simulate it with a loudspeaker 1 meter away playing pre-recorded speech at 3 dB SNR, matching the user’s voice volume. This poses a challenge for earables due to the similarity to the user’s speech. Spectrograms (Fig. \ref{fig:nospeaker}, \ref{fig:speaker}) show that microphone audio is distorted by the competing speaker (highlighted in Fig. \ref{fig:speaker}), while accelerometer data, capturing bone-conducted vibrations, remains unaffected, attenuating high frequencies. Thus, accelerometers effectively isolate the user’s voice from environmental noise, enhancing speech clarity.

% \noindent\textbf{Competing speaker} refers to the scenario when there is another speaker around the user. We use a loudspeaker positioned one meter away to mimic it by playing a pre-recorded speech.
% The volume of the competing speaker was set to match the user’s voice, with a signal-to-noise ratio (SNR) of 3 dB. 
% The competing speaker represents a significant challenge for earable devices, as it is similar to the user's speech.
% In Fig. \ref{fig:nospeaker} and \ref{fig:speaker} where we present the spectrograms of microphone audio and accelerometer data in either a quiet environment or with a competing speaker. We observe that the accelerometer captures acoustic vibrations that attenuate during bone conduction, particularly at higher frequencies. 
% Otherwise, the microphone audio spectrogram exhibits significant distortions from competing speakers, as highlighted by the white box in Figure \ref{fig:speaker}. In contrast, the accelerometer spectrograms remain unaffected, as the competing speaker’s sound does not produce bone-conducted vibrations.
% As a result, the accelerometer effectively isolates the user's voice from environmental sounds transmitted through the air, making it highly suitable for speech enhancement applications.

\noindent\textbf{User motion} may impact the reading of the accelerometer. In Figure \ref{fig:walk}, we illustrate that walking introduces low-frequency noise in the accelerometer data. The green boxes highlight a zoomed-in view of signals below 100 Hz, revealing clear periodic fluctuations at low frequencies (i.e., $<50 Hz$), which correspond to the volunteer’s steps. Consequently, user motion primarily affects the accelerometer at lower frequencies, leaving speech on higher frequencies unaffected.
% Additionally, we observed subtle low-frequency fluctuations in the accelerometer data in Fig. \ref{fig:competing_speaker}, which can be effectively removed by a high-pass filter since human speech typically occurs above 85 Hz \cite{Voicefre75:online}.

\noindent\textbf{Frequency Response} is the key property of BCF, as we can observe in Fig. \ref{fig:nospeaker} and \ref{fig:speaker} where the vibration signal lacks high-frequency components, consistent with the low-pass filter effect described earlier. We leave the details of the modeling of BCF for the later section.

\begin{figure}
   \begin{minipage}[t]{0.45\columnwidth}
     \centering
    \includegraphics[width=1\textwidth]{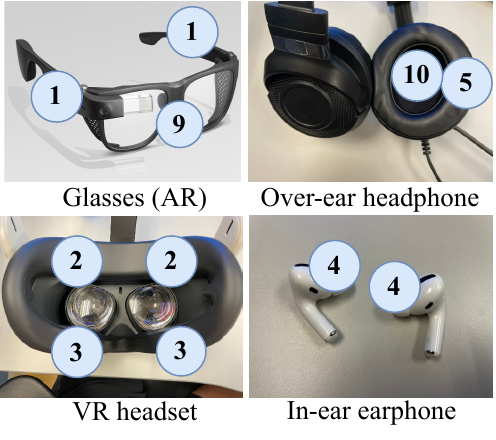}
    \caption{Potential placements on devices.}    
    \vspace{-1em}
    \label{fig:devices}
    \end{minipage}
    \hfill
\begin{minipage}[t]{0.45\columnwidth}
     \centering
    \includegraphics[width=1\textwidth]{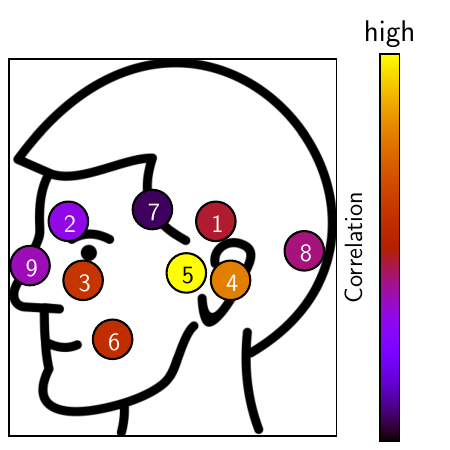}
    \caption{Intensities of received bone-conducted vibrations on ten locations of the head.}
    \vspace{-1em}
    \label{fig:head_positison}
\end{minipage}
\end{figure}

\noindent\textbf{Locations on the head} can impact the BCF property since it indicates a different propagation path. We measured bone-conducted vibration intensities at ten head locations (Fig. \ref{fig:head_positison}): upper ear, eyebrow, cheekbones, ear, temporomandibular joint, cheek, temple, back of the head, nose, and headphone (EarSense taped to an over-ear headphone pad). Some locations suit commercial earables like glasses and headsets (Fig. \ref{fig:devices}). Pearson correlation coefficients between audio and acceleration, visualized in Fig. \ref{fig:head_positison}, show vibrations at most locations, with higher correlations near the mouth, enhancing clean speech extraction.
% \noindent\textbf{Locations on the head} can impact the BCF property since it indicates different propagation path.
% As shown in \fig\ref{fig:head_positison}, we select ten unique locations of interest on the head and measure the intensities of received bone-conducted vibrations: \#1 \emph{upper ear}, \#2 \emph{eyebrow}, \#3 \emph{cheekbones}, \#4 \emph{ear}, \#5 \emph{temporomandibular joint}, \#6 \emph{cheek}, \#7 \emph{temple}, \#8 \emph{back of the head}, \#9 \emph{nose} and \#10 \emph{headphone}, which refers to taping EarSense on the interior of the pad of an over-ear headphone. Among those positions, some are compatible with commercial earables like glasses, headphones, and VR headsets as illustrated in \fig\ref{fig:devices} corresponding to the number in \fig\ref{fig:head_positison}. 
% We compute Pearson correlation coefficients between the audio and the acceleration on each location and mark the values using a color map in \fig\ref{fig:head_positison}.
% We observe that bone-conducted vibrations exist at most locations of the head. In particular, locations closer to the mouth show higher correlations, indicating a larger possibility of extracting clean speech. 

\noindent\textbf{Summary:} a) Bone-conducted vibration is robust to environmental voice, b) The user's motion only generates vibrations lower than $85\,\text{Hz}$, c) BCF has a unique frequency response, and d) We can receive bone-conducted vibration from multiple locations on the head.
\section{Methodology}
\subsection{Overview of \systemname{}}
\label{sec:overview}

\begin{figure}
  \centering
  \includegraphics[width=1\linewidth]{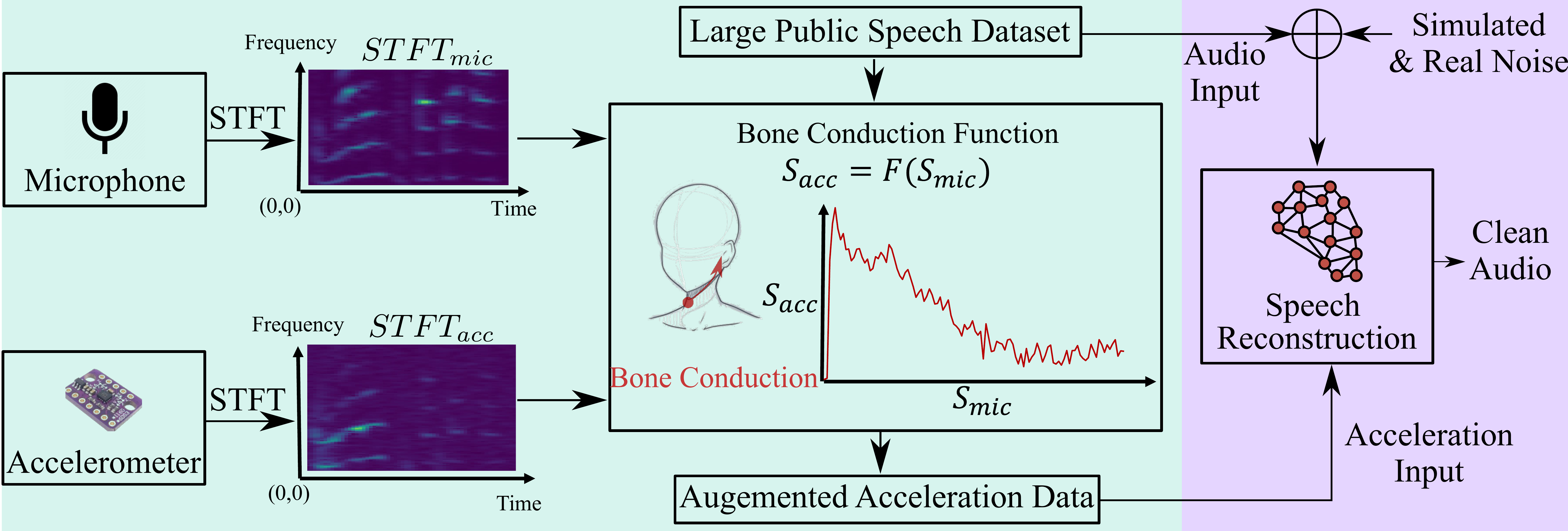}
  \caption{The overview of \systemname{} system.}
  \label{fig:overview}
    \vspace{-1em}
\end{figure}

We introduce \systemname{}, a novel speech enhancement system designed for earables, as illustrated in \fig\ref{fig:overview}. The system comprises three core components: a multi-modal speech enhancement network, pre-training with BCFs (BCFs), and adaptation strategies for speech enhancement.

First, the multi-modal speech enhancement network leverages the complementary strengths of audio and vibration signals, as detailed in \sect\ref{sec:network}. Unlike traditional methods that rely on time-invariant speaker embeddings, \systemname{} employs time-dependent features from vibration signals to preserve fine-grained information, balancing compression and performance for robust speech extraction.

Second, to address the scarcity of paired audio-vibration datasets, we propose a pre-training strategy using BCFs, as described in \sect\ref{sec:bcf}. By estimating BCFs from limited recordings and applying them to public audio datasets like LibriSpeech, we generate synthetic vibration data for training. This approach ensures diverse audio-vibration pairs, enhancing the model's generalization across users and conditions.

\new{
Lastly, \systemname{} is expected to perform reliably in diverse, noisy environments where clean paired data is scarce. Traditional supervised learning struggles in such conditions, requiring robust algorithms adaptable to varied noises. We propose an adaptive speech enhancement that boost \systemname{} in both training and testing. For the adaptive training, we propose a continual self-supervised learning framework (\sect\ref{sec:continual_learning}) that leverages noisy audio via multi-modal SNR estimation and SNR-aware training, enabling effective use of noisy data for robust performance. For the adaptive testing, we propose an adaptive inference framework (\sect\ref{sec:adaptive_inference}) that adjusts computational resources based on estimated noise levels, modulating separator block depth to balance quality and efficiency. We re-trained the speech enhancement model by multi-level loss function, enabling arbitrary number of blocks conditioned on the noise level for resource-constrained devices.
}

\subsection{Multi-Modal Speech Enhancement}
\label{sec:network}

\subsubsection{Problem formulation}
Suppose the recordings of the microphone and vibration sensor (e.g., accelerometer) are $S_m$ and $S_v$, respectively. Generally speaking, our design goal is to extract the desired conversation from a mixture of speech under the vibration sensor as a condition. Compared to other conditional speech extraction models like voicefilter \cite{wang2018voicefilter}, the vibration provides fine-grained information but lacks the high-frequency component, as illustrated before.

\subsubsection{Model architecture}

\begin{figure}
    \centering
    \includegraphics[width=1\linewidth]{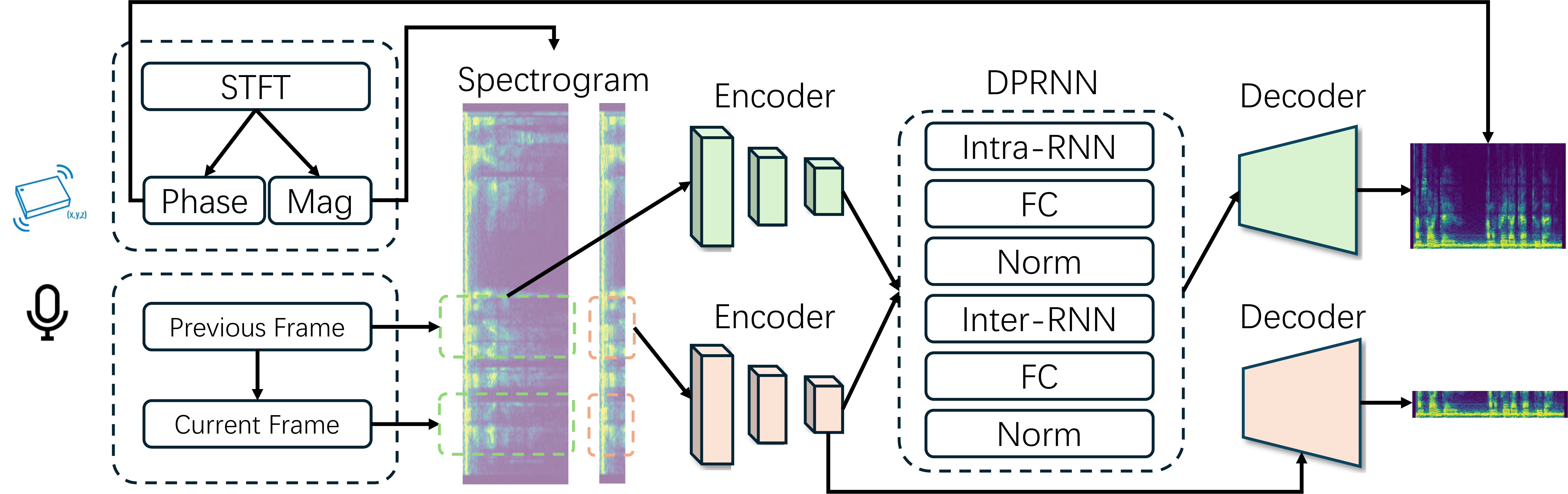}
    \caption{Network architecture of multi-modal speech enhancement network.}
    \label{fig:network}
        \vspace{-1em}
\end{figure}

We illustrate the neural network architecture as follows, which is shown in Fig. \ref{fig:network}.

\noindent\textbf{Features}.
Our speech enhancement approach operates in the time-frequency domain. We transform both waveform audio and vibration signals into spectrograms using the Short-Time Fourier Transform (STFT). Note that the three axes acceleration is first normalized (L2) and converted to a spectrogram. The STFT output is separated into phase and magnitude components, with our primary neural network processing only the magnitude.
Due to the typically higher sampling rate of audio compared to vibration, we adjust the STFT parameters accordingly to make both of them have the same time dimension. For instance, we set the window size for audio to 640, while the vibration window size is set to 64 to account for the difference. 

\noindent\textbf{Encoder}.
We employ convolutional neural networks (CNNs) to extract high-level features from the two modalities, respectively. The features from both encoders are concatenated along the channel dimension. Each encoder is constructed by stacking basic blocks, where each block comprises a 2D convolutional layer, batch normalization, ReLU activation, and max-pooling. To accelerate training, we incorporate a residual shortcut from the block input to the layer before the final deconvolution. To enhance the receptive field and capture harmonic patterns across the entire spectrogram, we use dilated convolutions instead of standard ones.
The audio data has a sampling rate of 16 kHz, which is ten times higher than the acceleration data sampling rate of 1.6 kHz. As a result, we use different window lengths for the audio and vibration, which leads to a ten times larger frequency dimension than that of the vibration data. To merge the differences, we apply three more layers to the audio branch and a projection layer at the end of the vibration encoder to ensure that the dimensions are aligned.

\noindent\textbf{DPRNN}.
After the encoder, we aim to separate speech and noise at the feature level, drawing on concepts from sound separation models. To achieve this, we incorporate the DPRNN \cite{luo2020dual} neural network, known for its lightweight and effective performance in sound separation. Specifically, DPRNN employs two RNNs: one for inter-block modeling (time dimension) and another for intra-block modeling (frequency dimension). We configure DPRNN to focus on a single source, as our goal is to isolate speech without estimating both noise and speech.

\noindent\textbf{Decoder}.
We designed two decoders: a fusion decoder and an auxiliary decoder. Both share the same block structure as the encoder but feature a decreasing number of filters and increasing output tensor sizes. The fusion decoder processes concatenated features from both modalities to generate a spectrogram mask, which is applied to the original noisy audio spectrogram through element-wise multiplication to produce a clean spectrogram. We then incorporate the noisy phase and apply inverse STFT to reconstruct the waveform audio. The auxiliary decoder, which processes only accelerometer features, predicts the low-frequency component of the clean audio. We add the auxiliary decoder since the audio branch contains much more information than the vibration branch. During training, the model may discard the vibration branch since audio-only speech enhancement is also valid when the noise is different from the target speech. To compensate for the collapse, forcing the vibration branch to reconstruct itself can make sure the features are valid.

\noindent\textbf{Real-time inference}.
Real-time speech enhancement is critical to minimize user experience disruptions. When processing audio at the frame level, the processing time per frame must be shorter than the frame duration to ensure the end-to-end latency equals the frame length.
However, models like encoders, decoders, and separators typically depend on the current frame, past frames, and sometimes future frames. To enable frame-by-frame inference, previous hidden states must be retained and utilized in subsequent frames. To further reduce latency, all convolutional layers should employ causal convolution, which relies solely on past frames. Additionally, the RNN module in DPRNN should be configured as unidirectional to support streaming processing.

\noindent\textbf{Loss}.
The fusion decoder uses the SISNR loss, where we use $s$ and $\hat{s}$ to represent the clean and enhanced signals as follows: 
\begin{align} 
\label{math:sisnr_loss}
\begin{split}
    L = 20 \log_{10} \left( \frac{\| s \|^2}{\| s - \hat{s} \|^2} \right)
\end{split}
\end{align}
The above loss applies to full-band audio, which may omit the vibration data that only exists in the low-band. Consequently, we have another auxiliary loss that applies to the extra decoder for the vibration data. Different from the above loss, we use the low-passed clean audio as the target instead. We set a weight of 0.05 for the auxiliary loss to balance the scales of the two losses.

\subsection{Pre-training with BCF}
\label{sec:bcf}

\begin{figure}[ht]
    \centering
    \begin{minipage}{0.48\linewidth}
        \centering
        \includegraphics[width=1\linewidth]{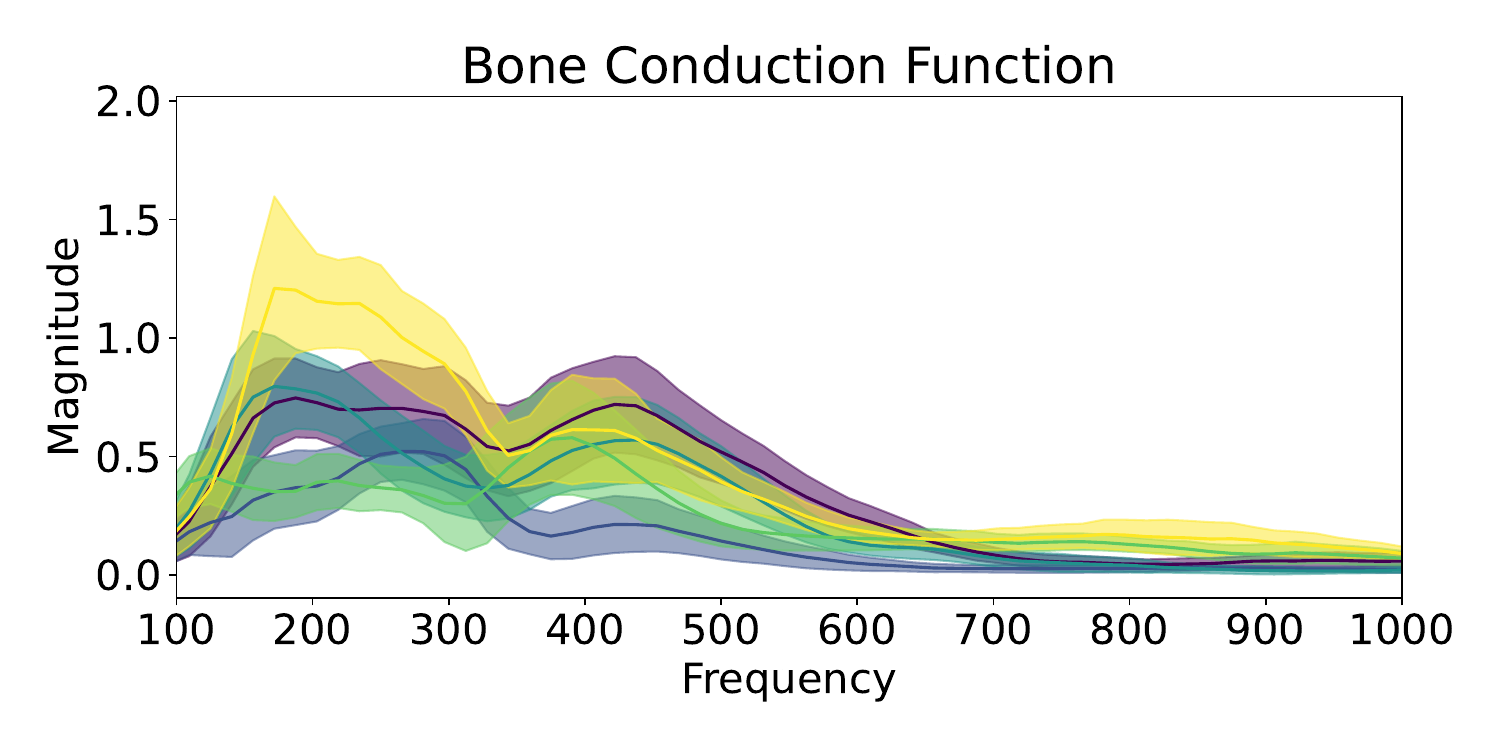}
    \caption{BCF of dataset 1.}    \vspace{-1em}

    \label{fig:bcf_ABCS}
    \end{minipage}
    \hfill
    \begin{minipage}{0.48\linewidth}
        \centering
       \includegraphics[width=1\linewidth]{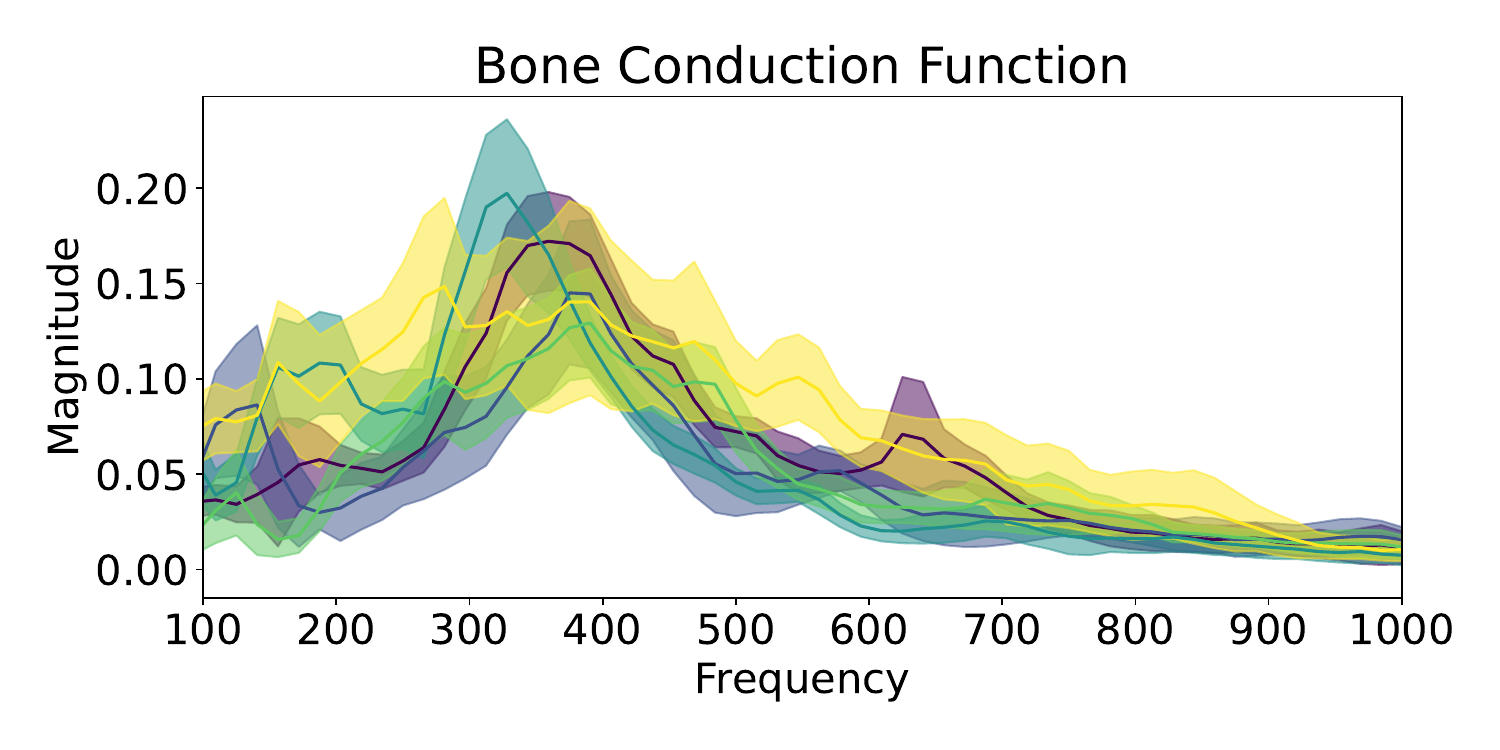}
    \caption{BCF of dataset 2.}    \vspace{-1em}

    \label{fig:bcf_EMSB}
    \end{minipage}
\end{figure}

\subsubsection{Problem formulation}
Motivated by our findings in \sect\ref{sec:measurement}, bone-conducted vibration is a promising complementary sensing modality to microphone recording for speech enhancement under environmental noises. 
However, there is no large dataset with paired acceleration and audio on earabless available.
As a result, we consider using the BCF, which depicts the transfer function from audio to vibration, to generate a virtual vibration dataset. We note that this estimation does not resort to black-box deep learning approaches, which require a huge amount of training data. Instead, we take advantage of prior knowledge that a frequency response exists between the acceleration and audio spectrograms.  

\subsubsection{Function estimation}
\label{sec:function}
To estimate the BCF, we split the paired data into 5-second windows, and each window can contribute one sample of BCF. Specifically, the power spectral density (PSD) is considered the frequency response between the two signals, which can be estimated using Welch’s method \cite{welch2003use}. This method involves dividing the data into overlapping segments, computing a modified periodogram for each segment, and then averaging these periodograms. Since both audio and vibration signals are sparse in the frequency domain, the estimated PSD may not always be reliable. To address this, we model the BCF as a Gaussian distribution in the frequency domain because it exhibits and similar pattern with non-trivial variance due to the complex structure of the head skeleton \cite{chang2016development}, as illustrated in Fig. \ref{fig:bcf_ABCS} and \ref{fig:bcf_EMSB}. Specifically, we denote this function as $f\sim N(\mu,\sigma^2)$, in which $\mu$ and variance $\sigma$ contribute to the contour and fluctuation of frequency response, estimated from the recording of a group of users.

\begin{figure}
    \centering
    \includegraphics[width=0.75\linewidth]{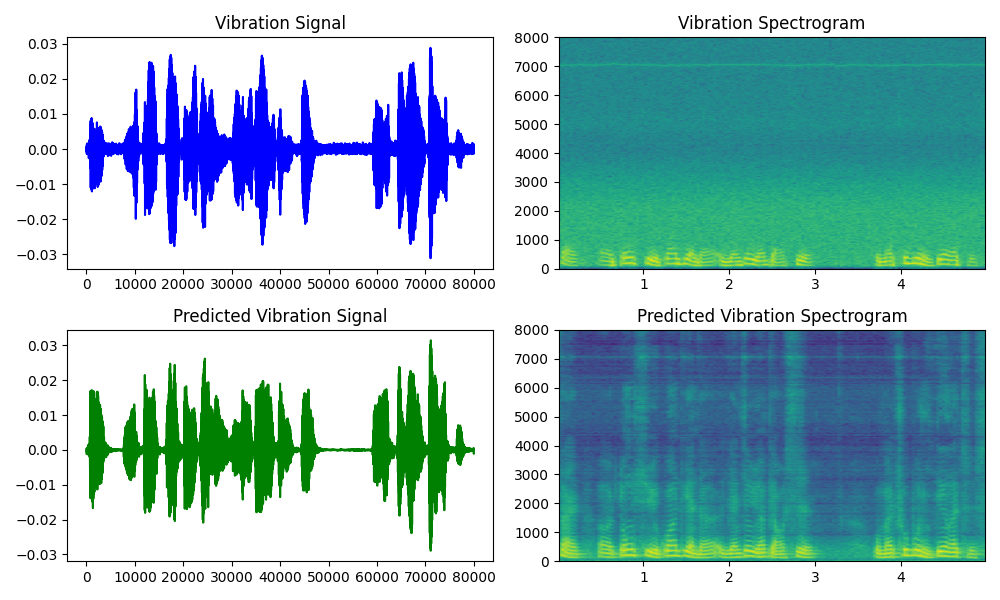}
    \caption{Vibration and the reconstructed vibration.}
    \label{fig:synthetic_compare}
    \vspace{-2em}
\end{figure}

\subsubsection{Data Augmentation with BCFs}
\label{sec:aug}
We develop a data augmentation approach with BCFs described in \sect\ref{sec:function}.
Note that we cannot apply the inverse BCF to turn the acceleration back to audio due to the significantly limited sample rate of the acceleration data. In addition, the frequency band (larger than $500\,\text{Hz}$) has very slim energy, which can cause a large energy after the inversion.
We utilize these functions to generate acceleration signals using a large-scale audio dataset, i.e., LibriSpeech \cite{panayotov2015librispeech}. 
To be specific, for each audio clip, we first select a BCF (i.e., a list of means and variances over different frequencies) from the pool randomly. 
Then, we restore the frequency response from the Gaussian distribution of given parameters. Lastly, we can augment the audio with synthetic acceleration data by directly multiplying the frequency response. 
Fig. \ref{fig:synthetic_compare} shows the spectrograms of augmented acceleration and real acceleration signals, respectively. The augmented spectrogram is close to the real acceleration spectrogram. 
We compute the similarity by calculating the mean of the absolute distance of all pixels in the whole spectrogram and dividing it by the largest value of the real acceleration spectrogram. The average error for all volunteers is only 4.5\%, which indicates that our proposed acceleration augmentation is reliable and reduce the data collection overhead.

% Since the BCF depends on both the user's position and individual characteristics, the BCF derived from our self-collected dataset may not align with the BCF in our test set. Therefore, we need to implement a fine-tuning stage for each user to enhance performance after pre-training with the synthetic dataset. A more effective pre-training process will reduce the effort required for fine-tuning, making it more user-friendly.
% To achieve this, we generate a pool of BCFs by applying cubic interpolation to the existing functions in our dataset. Additionally, we utilize a Gaussian distribution based on previously estimated parameters to minimize overfitting in our model and improve its generalizability across different users.
\new{
\subsection{Adaptive Speech Enhancement}

\subsubsection{Problem formulation}
In speech enhancement, performance depends not only on the quality of bone-conducted vibrations—which are influenced by personal physiological factors and hardware quality, as discussed earlier—but also on the characteristics of interfering noise. For instance, louder noise typically presents a greater challenge for suppression.

To improve the performance of \systemname{}, we propose making the system noise-aware, allowing for adaptive speech enhancement. By leveraging noise characteristics, we can dynamically adjust enhancement strategies, unlocking new capabilities in real-world scenarios. This approach involves two key components:
\begin{itemize}
    \item Noise-aware training – Training a model exclusively on noisy data to improve robustness.
    \item Noise-aware inference – Optimizing inference under limited computational resources while maintaining performance.
\end{itemize}
Before implementing these strategies, however, we must first accurately estimate the noise profile.

\subsubsection{Noise estimation}
\label{sec:noise_estimation}

Based on the analysis above, it is essential to first obtain knowledge of noise. 
We conclude that there are two impact factors of noise: noise type and noise levels. The former can be represented by the audio classification of the noise, which is relatively mature \cite{schmid2023efficient}. On the other hand, the noise level can be represented by the SNR.

We have observed that there is a strong correlation between bone-conducted vibration and the user's speech. The above property also indicates that the combination of audio and vibration can infer the SNR. Specifically, the correlation between audio and vibration decreases when the audio is contaminated by noise. Therefore, we propose a multi-modal SNR estimator that builds upon the audio-only SNR estimator described in \cite{subakan2021realm}.
Specifically, we transform both audio and vibration by STFT and keep the magnitude only. We conduct zero-padding for the vibration spectrogram is necessary. Then, the spectrograms are concatenated.
The estimator consists of five convolutional layers, each with a kernel size of 4, 128 channels, and a stride of 1. This is followed by a statistical pooling layer and two fully connected layers, each containing 256 neurons. We use ReLU (Rectified Linear Unit) activation functions for all the layers.
To ensure the estimated SI-SNR values range between -20 and 20 dB, we normalize the network's output to fall between 0 and 1. In this normalization, a value of 0 corresponds to -20 dB, while a value of 1 corresponds to 20 dB. This range compression is achieved using a sigmoid function applied to the output of the network.

\subsubsection{Noise-aware training}
\label{sec:continual_learning}
\begin{figure}
  \centering
  \includegraphics[width=1\linewidth]{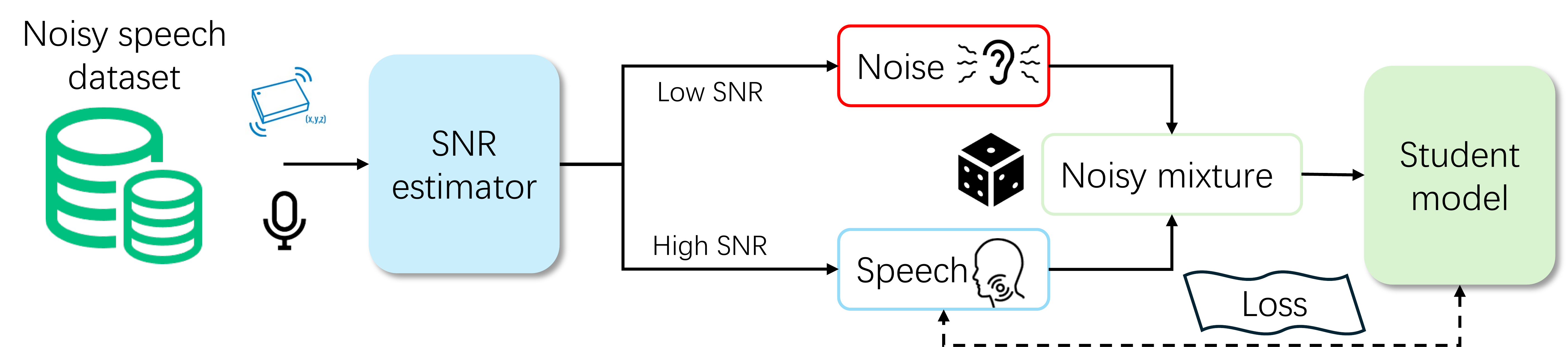}
  \caption{Continual learning (adaptive training) of \systemname{}.}
  \label{fig:continual_learning}
      \vspace{-1em}
\end{figure}

Given the obtained noise information—including noise type and noise level—we first focus on addressing noise type. Similar to target speech enhancement \cite{veluri2024look} or target sound enhancement \cite{veluri2023semantic}, conditioning the enhancement process on noise type is a promising approach. However, noise type is highly dynamic and cannot be sufficiently represented by either a single class or a static feature. Instead, we characterize noise type through the concept of noise domain, which captures the distribution of noise sources that vary spatiotemporally (e.g., by location and time).
To explore this, we trained our multi-modal network using in-domain noise samples: Mandarin speech from AI-SHELL \cite{bu2017aishell} and general environmental noise from FSD50K \cite{fonseca2021fsd50k}. When evaluating the model on out-of-domain noise—such as English speech (TIMIT \cite{fisher1986darpa}), self-noise from the speaker, or environmental sounds (DEMAND \cite{thiemann2013demand})—we observed a significant performance degradation, as shown in Fig. \ref{fig:cross_noise}. The red line (in-domain performance) consistently surpasses results for English noise and self-noise, highlighting the challenge of domain generalization.

Consequently, to enable \systemname{}'s capability to work on the noise type, equivalent to domain adaptation to a new noise domain in a continuous manner. 
A straightforward solution is to fine-tune the model with out-of-domain noise, effectively making it in-domain.
Considering it is not practical to ask the user to record clean speech and noise as we do in the lab, self-supervised or unsupervised learning becomes necessary.
Specifically, our problem can be formulated as an unsupervised continual learning where we only have access to noisy audio that contains in-domain noise.
As a naive solution, we can consider the noisy audio as the clean audio; there is no doubt that it can not perform as well as clean audio, or even collapse when the data is extremely noisy. However, we observe that there are also relatively clean audio samples in the noisy data, so it is critical to find out those samples (high-quality data) effectively. Suppose we can select the high-quality data, then we also need an effective algorithm to utilize it.

Suppose the noisy audio is $S_m$ and vibration is $S_v$, the estimated SNR is $E = Estimator(S_m, S_v)$. In our initial approach, we only select samples with an estimated SNR above a certain threshold, categorizing these as positive samples. Conversely, samples with an estimated SNR below another threshold are classified as negative samples. We can then mix these positive and negative samples offline to fine-tune the model.
While this method is effective, it discards a large portion of the data, limiting performance due to the reduced availability of high-quality samples. To enhance the quality of our dataset, we can preprocess both the noisy audio and vibration using a pretrained speech enhancement model.
The pretrained model, which plays a similar role as the teacher model in RemixIT \cite{tzinis2022remixit}, can be updated at the end of each epoch to ensure continuous enhancement of our data quality.

\begin{algorithm}
\caption{Continual learning for the noisy dataset \( D_m \)}
\begin{algorithmic}[1]
\State \(\{\mathbf{D}_{\text{clean}}, \mathbf{D}_{\text{noise}}, \mathbf{D}_{\text{noisy}}\} = \{\}, threshold = \beta\)

\For {\(\text{each batch } \mathbf{m} \in \mathcal{D}_m, \mathbf{m} \in \mathbf{R}^{B \times T}\)}
    \State \(\text{SNR} \gets \text{ESTIMATE\_SNR}(\mathbf{m})\) 

    \If {$\lvert SNR \rvert >threshold$}
         \State \(\{\mathbf{D}_{\text{clean}}, \mathbf{D}_{\text{noise}}\} \gets (\mathbf{m}, \text{SNR})\) 
    \EndIf
    
    \If {\(\mathcal{D}_{\text{clean}}, \mathcal{D}_{\text{noise}}\) is enough}

    \State \((\mathbf{D}_{\text{noisy}})\ \gets \text{REMIX\_DATASET}(\mathbf{D}_{\text{clean}}, \mathbf{D}_{\text{noise}})\)
    
    \EndIf
    
\EndFor
\end{algorithmic}
\end{algorithm}

\subsubsection{Noise-aware inference}
\label{sec:adaptive_inference}

\begin{figure}
  \centering
  \includegraphics[width=1\linewidth]{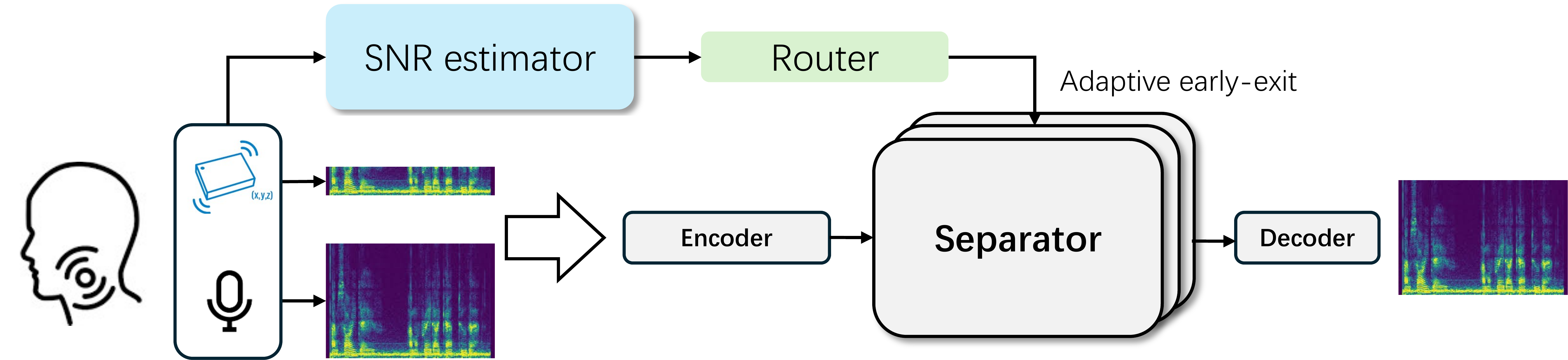}
  \caption{Adaptive inference of \systemname{}.}
  \label{fig:adaptive}    
  \vspace{-1em}
\end{figure}

Beyond noise type, noise level also critically impacts speech enhancement—but in a distinct way. Unlike noise type, it is uncommon for a model to perform well only on strong noise while failing on weak noise. Instead, noise level exhibits a clear correlation with task difficulty: higher noise levels generally demand more aggressive enhancement, while lower levels may require minimal processing.
This property opens opportunities to optimize inference efficiency. During deployment of our multi-modal speech enhancement system, we observed dynamic variations in both user speech presence and noise intensity. Running the same enhancement model continuously—even on relatively clean signals—wastes computational resources and introduces unnecessary latency. A straightforward optimization is to integrate voice activity detection (VAD) using head vibration signals \cite{schilk2023ear}, activating enhancement only when speech is detected.

However, even during active speech, noise levels can fluctuate substantially (e.g., from loud to near-absent), making on-off enhancement inefficient. To address this, we propose an adaptive \systemname{} that dynamically adjusts its processing based on input characteristics. The key is identifying a conditioning factor that reliably reflects enhancement difficulty (e.g., real-time noise level). By feeding this factor to the model as an input, we can selectively allocate computational resources—minimizing overhead without sacrificing performance.

Based on prior analysis, noise level serves as the key factor for enabling adaptive speech enhancement by leveraging the previously proposed SNR estimator. Specifically, we can set an SNR threshold and dynamically adjust the depth of the speech enhancement model until the output meets the desired threshold. The number of separator blocks is an ideal control factor, as it significantly impacts computation without altering the model’s pipeline. To support adaptive inference, the training process for speech enhancement must also be modified. This involves defining multiple loss functions corresponding to different numbers of modules, which are averaged to compute the final loss as follows:
\[
L = \sum_{i=0}^{N} w_{i} 20 \log_{10} \left( \frac{\| s \|^2}{\| s - \hat{s}_i \|^2} \right)
\], where $\hat{s}_i$ refers to the output after the $i^{th}$ separator modules and $w_{i}$ refers to the weight.
}

\section{Experiment Setup}

\subsection{Dataset}
\begin{table}
    \centering
    \begin{tabular}{|c|c|c|c|}
    \hline
       Dataset name  &  Content & Duration & Role\\
       \hline
       LibriSpeech-train  & English & 1000 hours&  Pre-train\\
        \hline
       LibriSpeech-dev  & English & 1000 hours & Noise\\
       \hline
       Ai-shell & Mandarin &  hours & Noise \\
        \hline
        VibVoice & English & 3 hours  & Fine-tune\\
       \hline
       FSD50K & General sound &  hours & Noise \\
       \hline
       % EMSB  & Mandarin & 128 hours & 287 & Fine-tune \\
       %  \hline
       % ABCS  & Mandarin & 42 hours & 100 & Fine-tune\\
       %  \hline
     \end{tabular}
    \caption{Information about the dataset used.}
    \label{tab:my_label}
        \vspace{-1em}

\end{table}

For pre-training with Bone Conduction Functions (BCFs) in Sec. \ref{sec:bcf}, we use LibriSpeech \cite{panayotov2015librispeech} as the source dataset. Besides, we recruited 15 volunteers to collect audio-vibration datasets in a clean lab and noisy real-world settings, reading English content from LibriSpeech \cite{panayotov2015librispeech}. Each volunteer contributed 10 minutes of data.
Noise Datasets are assumed to be clean by default. For training and evaluation, we add noise with SNRs from -5 dB to 15 dB (average input SNR of 5 dB). Noise types, in equal proportions, include: 1) FSD50K \cite{fonseca2021fsd50k} general noise, 2) speech noise from Ai-shell \cite{bu2017aishell} or LibriSpeech \cite{panayotov2015librispeech}, and 3) self-noise from the same user (different utterance). Audio is convolved with room impulse responses from \cite{hadad2014multichannel} to mimic real-world conditions.

\subsection{Metrics}
% Defining Signal-to-Noise Ratio (SNR)
\noindent\textbf{Signal-to-Noise Ratio (SNR)}: Measures signal quality relative to noise, computed as
\(\text{SNR}(x, y) = 20 \log_{10} \left( \left( \frac{y}{x - y} \right)^2 \right),\)
where $x$ is the estimated audio and $y$ is the clean audio. Higher values indicate better quality. Scale-invariant SNR is used by default.
% Defining Perceptual Evaluation of Speech Quality (PESQ)
\textbf{Perceptual Evaluation of Speech Quality (PESQ)}: Per ITU-T P.862, assesses speech quality with scores from 1 (poor) to 5 (excellent). Wide-band version evaluates full-band speech.
% Defining Log-Spectral Distance (LSD)
\textbf{Log-Spectral Distance (LSD)}: Measures frequency-domain quality between reconstructed and ground truth audio:
\(\text{LSD}(x, y) = \frac{1}{L} \sum_{l=1}^{L} \sqrt{\frac{1}{K} \sum_{k=1}^{K} \left( X(l,k) - \hat{X}(l,k) \right)^2},\)
where $l$ and $k$ are time and frequency indices, $X = \log(| \text{STFT}(y) |^2)$, and $\hat{X} = \log(| \text{STFT}(x) |^2)$. Lower values indicate higher quality.

\subsection{Baselines}
We deploy two baselines, i.e., FullSubNet (FSN) \cite{hao2021fullsubnet} and SEANet (SN) \cite{tagliasacchi2020seanet}.
FSN and SN are two state-of-the-art speech enhancement approaches using audio-only and audio-vibration inputs, respectively.
We train SN using our dataset since the one used in its paper is not available to the public. Specifically, we train SN using vibration data with a sample rate of 1.6kHz to ensure it is the same as \systemname{}.

\section{Evaluation}

\subsection{Overall Performance}
\label{sec:overall_eval}
\begin{figure}
    \begin{subfigure}{0.32\columnwidth}
        \centering
        \includegraphics[width=1\textwidth]{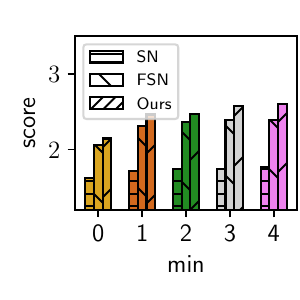}
        \vspace{-2em}
        \caption{PESQ.}
        \label{fig:calibration_0}
        \end{subfigure}
    \begin{subfigure}{0.32\columnwidth}
        \centering
        \includegraphics[width=1\textwidth]{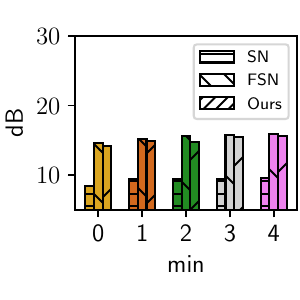}
        \vspace{-2em}
        \caption{SNR.}
        \label{fig:calibration_1}
         \end{subfigure}
    \begin{subfigure}{0.32\columnwidth}
        \centering
        \includegraphics[width=1\textwidth]{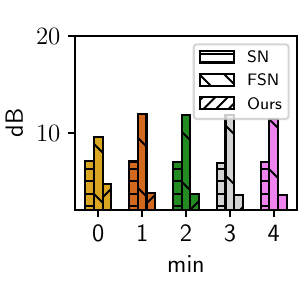}
        \vspace{-2em}
        \caption{LSD.}
        \label{fig:calibration_2}
         \end{subfigure}
    \caption{Impact of calibration time.}
    \label{fig:calibration}
    \vspace{-2em}
\end{figure}

\noindent\textbf{Calibration}.
\systemname{} operates out of the box but benefits from target-user data to enhance performance. Fig. \ref{fig:calibration} compares \systemname{} with two baselines using varying amounts of target-user data (zero indicates no user data during training). Results show that more calibration data improves performance across all methods, with \systemname{} outperforming baselines at equivalent data levels. \systemname{} achieves the highest PESQ for perceptual quality and the best LSD for spectrogram reconstruction, with SNR comparable to FSN due to similar time-domain signal outputs.

% \systemname{} can work out of the box, while data from the target user can further improve performance. \fig\ref{fig:noise_duration} shows the performance of \systemname{} and the two baselines with different amounts of data from the target user, in which zero means no target-user data is used during the training. The results show that longer target-user data can improve performance for all approaches,  and \systemname{} performs better than baselines with the same amount of calibration data. The calibration data can be collected continuously during usage without the user’s inputs/operations.
% \systemname{} achieves the best perceptual performance according to PESQ and the best reconstructed spectrogram according to LSD. \systemname{} and FSN have similar SNR results since they output similar signals in the time domain compared to the clean one. 
\begin{figure}
    \begin{subfigure}{0.32\columnwidth}
        \centering
        \includegraphics[width=1\textwidth]{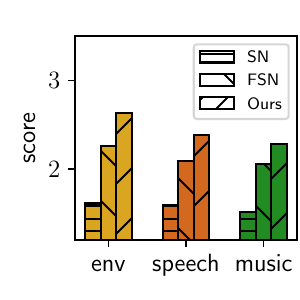}
        \vspace{-2em}
        \caption{PESQ.}
        \label{fig:noise_type_0}
        \end{subfigure}
    \begin{subfigure}{0.32\columnwidth}
        \centering
        \includegraphics[width=1\textwidth]{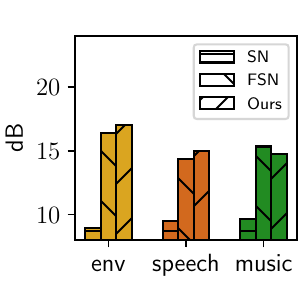}
        \vspace{-2em}
        \caption{SNR.}
        \label{fig:noise_type_1}
         \end{subfigure}
    \begin{subfigure}{0.32\columnwidth}
        \centering
        \includegraphics[width=1\textwidth]{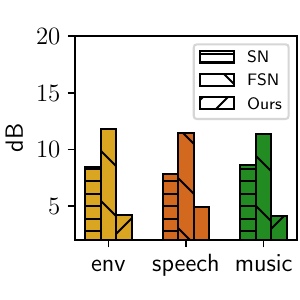}
        \vspace{-2em}
        \caption{LSD.}
        \label{fig:noise_type_2}
         \end{subfigure}

    \caption{Impact of noise types.}
    \label{fig:noise_type}
    \vspace{-2em}
\end{figure}

\noindent\textbf{Noise type}.
We evaluate the impact of different types of noises in \fig\ref{fig:noise_type}, i.e., environmental noises, competing speakers, and music. The result shows that \systemname{} performs better under all noises and metrics except for the SNR with music noise.  This is because the complex and dynamic spectrum of the user’s speech and music’s vocals introduce minor fluctuations in high frequencies, reflecting large fluctuations in SNR.
\begin{figure}
    %\vspace{-0.5em}
    \begin{subfigure}{0.32\columnwidth}
        \centering
        \includegraphics[width=1\textwidth]{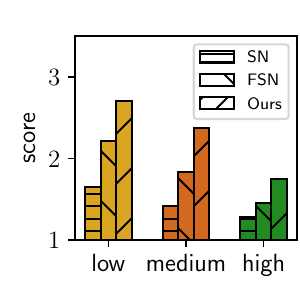}
        \vspace{-2em}
        \caption{PESQ.}
        \label{fig:noise_level_0}
        \end{subfigure}
    \begin{subfigure}{0.32\columnwidth}
        \centering
        \includegraphics[width=1\textwidth]{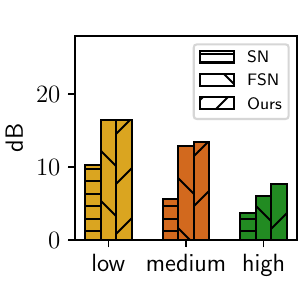}
        \vspace{-2em}
        \caption{SNR.}
        \label{fig:noise_level_1}
         \end{subfigure}
    \begin{subfigure}{0.32\columnwidth}
        \centering
        \includegraphics[width=1\textwidth]{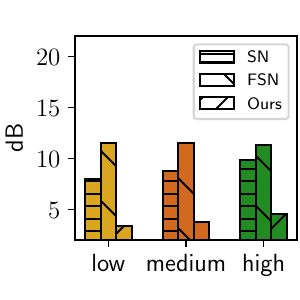}
        \vspace{-2em}
        \caption{LSD.}
        \label{fig:noise_level_2}
         \end{subfigure}

    \caption{Impact of noise levels.}
    \label{fig:noise_level}
    \vspace{-2em}
\end{figure}

\noindent\textbf{Noise level}.
We test \systemname{}'s performance under noise levels of low (10 dB), medium (5 dB), and high (0 dB with only speech noise). The results in \fig\ref{fig:noise_level} show that \systemname{} has better performance improvements, especially when the noise is challenging, i.e., 21$\%$ improvement on PESQ and 26$\%$ improvement on SNR. This is because the vibration can more robustly identify the target speech, whereas the audio-only solution is difficult to differentiate sound with a similar pattern (e.g., strong speech noise).
\begin{figure}
    \begin{subfigure}{0.32\columnwidth}
        \centering
        \includegraphics[width=1\textwidth]{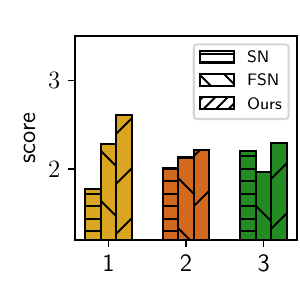}
        \vspace{-2em}
        \caption{PESQ.}
        \label{fig:noise_number_0}
        \end{subfigure}
        \hfill
    \begin{subfigure}{0.32\columnwidth}
        \centering
        \includegraphics[width=1\textwidth]{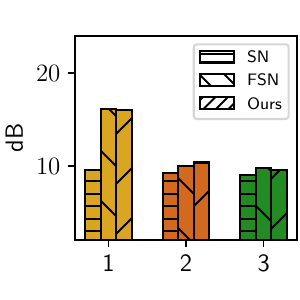}
        \vspace{-2em}
        \caption{SNR.}
        \label{fig:noise_number_1}
         \end{subfigure}
          \hfill
    \begin{subfigure}{0.32\columnwidth}
        \centering
        \includegraphics[width=1\textwidth]{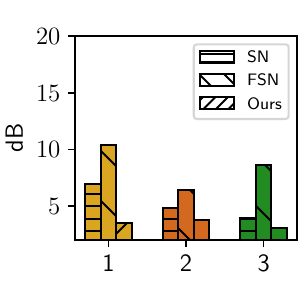}
        \vspace{-2em}
        \caption{LSD.}
        \label{fig:noise_number_2}
         \end{subfigure}

    \caption{Number of noise sources.}

    \label{fig:noise_number}
    \vspace{-2em}
\end{figure}

\noindent\textbf{The number of noise sources}.
We evaluate the impact of different numbers of noise sources by repeatedly mixing clean audio with random audio clips. \fig\ref{fig:noise_number} shows that \systemname{}'s performance degrades as the number of noise sources increases, but still outperforms all the baselines.

\noindent\textbf{Temporal stability}.
We further examine how \systemname{} performs for the same user over time. Note that the offset of sensor placement and minor changes in speech can cause a slight change. We collect ten-minute data from three volunteers twice, six months apart. 
 The results show that the performance of \systemname{} has negligible changes, from 2.6 to 2.5 for PESQ, 15.7 to 15.5 for SNR, and 4.3 to 4.6 for LSD. Besides, \systemname{} outperforms FSN, whose performance is 2.1 for PESQ, 15.2 for SNR, and 11 for LSD.
The results affirm that \systemname{} is robust to temporal changes.

\noindent\textbf{Airway blockage}.
Facial masks, which block air transmission, can reduce speech volume. We tested \systemname{} with three volunteers speaking identical content with and without masks. \systemname{}'s performance degrades minimally: 0.05 ($<2\%$) for PESQ, 0.4 ($<3\%$) for SNR, and 0.1 ($<3\%$) for LSD with masks. Compared to FSN (PESQ: 2.15, SNR: 13.8, LSD: 10), \systemname{} shows superior resilience, as expected, due to its use of bone-conducted vibration, unaffected by air transmission.

\begin{figure}
    \begin{subfigure}{0.3\columnwidth}
        \centering
        \includegraphics[width=1\textwidth]{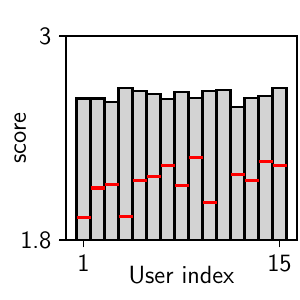}
        \vspace{-2em}
        \caption{PESQ.}
        \label{fig:per_user_0}
        \end{subfigure}
    \begin{subfigure}{0.3\columnwidth}
        \centering
        \includegraphics[width=1\textwidth]{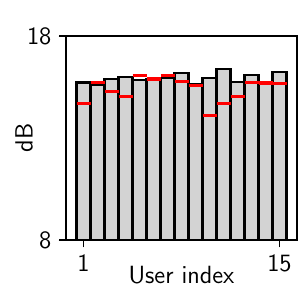}
        \vspace{-2em}
        \caption{SNR.}
         \end{subfigure}
        \label{fig:per_user_1}
    \begin{subfigure}{0.3\columnwidth}
        \centering
        \includegraphics[width=1\textwidth]{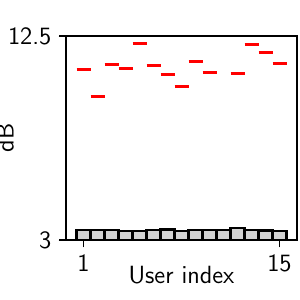}
        \vspace{-2em}
        \caption{LSD.}
        \label{fig:per_user_2}
         \end{subfigure}

    \caption{VibVoice on different users. Red line: the performance of the baseline, i.e., FullSubNet.}
    \label{fig:per_user}
    \vspace{-2em}

\end{figure}
\noindent\textbf{Variances among users}.
Speech and bone-conducted vibration can differ across users due to vocal features, head and skull shapes, body fat, etc.
\fig\ref{fig:per_user} shows the performance of \systemname{} across 15 users. \systemname{} shows stable and significantly better performance in PESQ compared to baseline and comparable performance in SNR.

\begin{figure}
    \begin{subfigure}{0.3\columnwidth}
        \centering
        \includegraphics[width=1\textwidth]{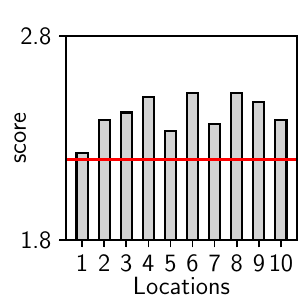}
        \vspace{-1em}
        \caption{PESQ.}
        \label{fig:location_0}
        \end{subfigure}
    \begin{subfigure}{0.3\columnwidth}
        \centering
        \includegraphics[width=1\textwidth]{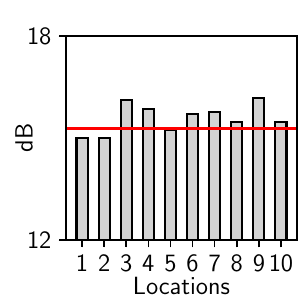}
        \vspace{-1em}
        \caption{SNR.}
        \label{fig:location_1}
         \end{subfigure}
    \begin{subfigure}{0.3\columnwidth}
        \centering
        \includegraphics[width=1\textwidth]{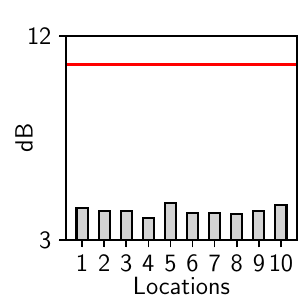}
        \vspace{-1em}
        \caption{LSD.}
        \label{fig:location_2}
         \end{subfigure}

    \caption{VibVoice on different head locations. Red line: the performance of the baseline, i.e., FullSubNet.}
    \label{fig:locations}
    \vspace{-1em}

\end{figure}

\noindent\textbf{Sensor positions}.
 We test \systemname{} when EarSense is placed in ten locations on the head as defined in \fig\ref{fig:head_positison}, validating \systemname{}'s effectiveness for different HMW devices.
The bars in \fig\ref{fig:locations} show \systemname{}'s performance at each location. The red line represents the performance of the baseline at \#4 \emph{ear}. 
The results show that \systemname{} achieves satisfactory performance at all locations in PESQ.
Note that locations like \#1 \emph{upper ear}, \#2 \emph{eyebrow}, \#5 \emph{temporomandibular joint}, \#7 \emph{temple}, and \#10 \emph{interior of the pad of the headphone} show similar or slightly lower SNR than the baseline, which is because the vibration intensity is slim due to their far distance to the audio source.

\noindent\textbf{Summary}.
 \systemname{} outperforms FSN and SN by up to $21\%$ for PESQ when the noise volume is low, where the PESQ for \systemname{} and FSN are 2.7 and 2.21, respectively. \systemname{} outperforms the baselines up to $26\%$ for SNR when the noise is speech with high volume, where the SNR of \systemname{} and FSN are 2.0 and 1.6, respectively. In addition, \systemname{} outperforms the baselines 50$\sim$80$\%$ in LSD under most impact factors, indicating the efficiency of our multi-modal design and novel data augmentation. \systemname{} has slightly lower performance in SNR for some cases, as it can be biased due to the similar spectrum of the user’s speech and music’s vocals. The dynamic also introduces minor fluctuations. In comparison, LSD evaluates the whole band without preference, so \systemname{} outperforms the two baselines by a large margin.

\subsection{Ablation Study}
\begin{table}
\begin{tabular}{cccc}\toprule
& PESQ&SNR&LSD \\ \midrule
VibVoice              & 2.6 & 15.6 & 3.5  \\ 
w/o auxiliary decoder & 2.5 & 15.1 & 4.4 \\ 
w/o augmentation      & 1.9 & 14   & 5\\ 
w/o Gaussian approx   & 2.4 & 15.2 & 4.2 \\ 
{\blue Accelerometer sample rate: 1200 Hz} & 2.47 & 14.4 & 4.2 \\ 
{\blue Accelerometer sample rate: 800 Hz}& 2.45 & 14.3 & 4.5 \\ 
{\blue Accelerometer sample rate: 400 Hz} & 2.4 & 14.2 & 4.4 \\ \bottomrule
\end{tabular}
\caption{\label{table3}Ablation study.}
\vspace{-2em}
\end{table}

We conduct an ablation study to understand the performance of different design components in \systemname{}. The performance of \systemname{} without different components is listed in Table \ref{table3}.

\noindent\textbf{No auxiliary decoder}. First, we remove the self-supervise loss, meaning the audio may dominate the model. The results indicate that the variant slightly degrades by 0.1 in PESQ, 0.3 in SNR, and 0.8 in LSD, respectively.

\begin{figure}
    \begin{subfigure}{0.32\columnwidth}
        \centering
        \includegraphics[width=1\textwidth]{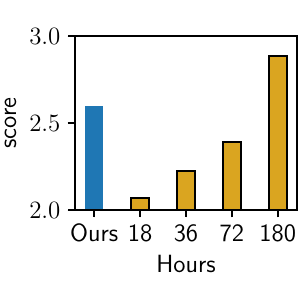}
        \vspace{-2em}
        \caption{PESQ.}
        \label{fig:data_augmentation_0}
        \end{subfigure}
    \begin{subfigure}{0.32\columnwidth}
        \centering
        \includegraphics[width=1\textwidth]{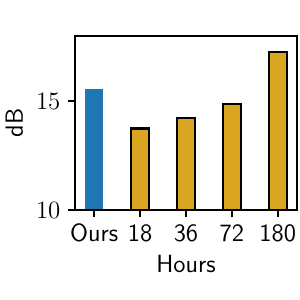}
        \vspace{-2em}
        \caption{SNR.}
        \label{fig:data_augmentation_1}
         \end{subfigure}
    \begin{subfigure}{0.32\columnwidth}
        \centering
        \includegraphics[width=1\textwidth]{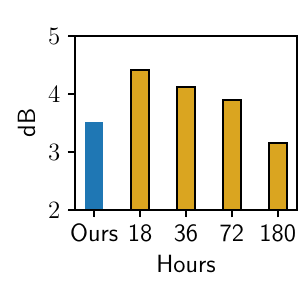}
        \vspace{-2em}
        \caption{LSD.}
        \label{fig:data_augmentation_2}
         \end{subfigure}
    \caption{Effectiveness of data augmentation. Blue bars: VibVoice using less than three hours of paired data with augmentation. Yellow bars: VibVoice using 18- to 180-hour paired data without augmentation.}
    \label{fig:data_augmentation}
    \vspace{-1em}
\end{figure}

\noindent\textbf{No data augmentation}.
Second, we remove the data augmentation based on the Bone Conduction Function. The performance significantly degrades to 1.9 in PESQ, 14 in SNR, and 5 in LSD. According to the definition of ITU and mean opinion score (MOS), the audio quality is poor when the score drops from 2.5 to 1.9 \cite{Perceptu66:online}. This confirms that small-scale self-collected datasets are insufficient for robust neural network training.
In addition, we compare \systemname{} trained on: a) 18–180 hours of paired audio-vibration data \cite{wang2022fusing} (5 kHz vibration bandwidth) and b) three hours of paired data with augmentation. Results in Fig. \ref{fig:data_augmentation} show that data augmentation achieves comparable performance with $\sim24\times$ less paired data.

\noindent\textbf{No Gaussian approximation}.
Third, the Bone Conduction Function is modeled by only the mean, while the variance is zero. The performance drops 0.2 in PESQ, 0.4 in SNR, and 0.5 for LSD, indicating that our Gaussian approximation is close to the nature of the Bone Conduction Function.

\noindent\textbf{Lower sample rate}.
 Lastly, we evaluate the performance of \systemname{} with a lower sample rate by downsampling the vibration data to 1200 Hz, 800 Hz, and 400 Hz. The results show that \systemname{} is robust to various sample rates, which consume less power in processing and communication. When the sample rate is 800 Hz, the output's PESQ only degrades $10\%$.

\subsection{Runtime Evaluation}
\label{sec:runtime}

\begin{table}
\centering
\begin{tabular}{cccc}\toprule
& \systemname{} & FSN & SN \\ \midrule
Desktop CPU    & 0.05 &0.27  & 0.5 \\ 
Desktop GPU    & 0.016&0.034 & 0.07\\ 
P30    & 0.16 &5     & 1.9 \\ 
Mate20 & 0.29 &4.6   & 1.7\\ 
Pixel7 & 0.31 &5.2   & 1.4\\ \bottomrule
\end{tabular}
\caption{\label{table4}Runtime analysis (second/instance).}
\vspace{-1.5em}
\end{table}

We test the execution latency of \systemname{} and the two baselines (i.e., FSN \cite{hao2021fullsubnet} and SN \cite{tagliasacchi2020seanet}) on a desktop PC (i.e., i7-11700k CPU and RTX 3060 GPU) and three smartphones (i.e., Huawei P30, Huawei Mate20, and Google Pixel 7). We run the inference of a 5-second clip 100 times and record the mean latency. 
The results in Table \ref{table4} show that \systemname{} reduces up to $31\times$ and $12\times$ less latency on average than FSN and SN, respectively. On the other hand, the results show that \systemname{} exhibits a greater advantage in runtime for low-end devices.
In conclusion, \systemname{} can support real-time voice applications with a delay of fewer than 0.6 seconds (i.e., two times the inference latency) for processing 5 5-second audio clip. The real-time factor is only 0.12, significantly less than the minimal requirement, which means less energy consumption and space to process other tasks.

\subsection{Extension Evaluation}

\begin{table}
\centering
\begin{tabular}{ccc}\toprule
SNR (dB)&ABCS&EMSB \\ \midrule
\systemname{} (TFGridNet) & 10.76 & 11.46 \\
\systemname{} (DPRNN) &7.96 &8.03  \\
TFGridNet & 9.25 & 10.54 \\
DPRNN & 3.04 & 3.08  \\
\bottomrule
\end{tabular}
\caption{Performance comparison on public datasets: ABCS and EMSB.}
\label{table5}
\vspace{-2em}
\end{table}

\begin{figure}
    \centering
    \begin{minipage}{0.48\linewidth}
        \centering
        \includegraphics[width=1\linewidth]{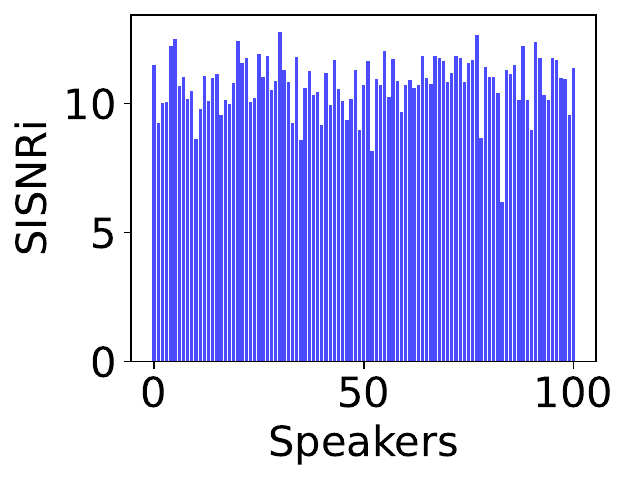}
        \caption{Performance for different speakers (ABCS).}
        \label{fig:eval_abcs}
    \end{minipage}\hfill
    \begin{minipage}{0.48\linewidth}
        \centering
        \includegraphics[width=1\linewidth]{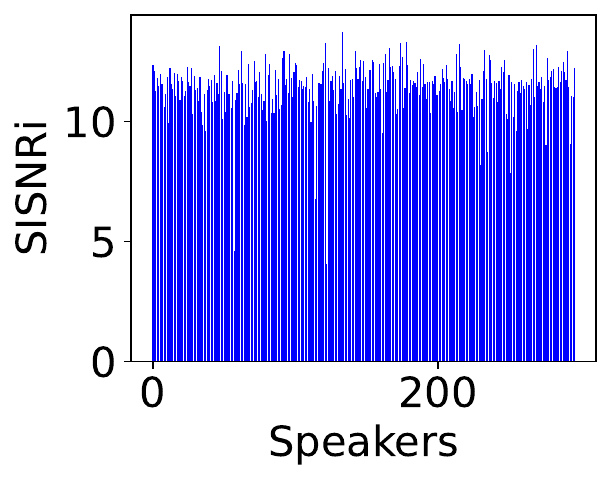}
        \caption{Performance for different speakers (EMSB).}
        \label{fig:eval_emsb}
    \end{minipage}
    \vspace{-1em}
\end{figure}

\new{
Except for the dataset collected from our prototype, we observe that there are a few public dataset that contains both vibration and synchronized audio, including EMSB \cite{wang2022fusing} and ABCS \cite{wang2022multi}. Specifically, both are collected in quiet places by customized earphones with a vibration sensor (without detailed documentation). Since both datasets are Mandarin, we consider the Mandarin dataset Ai-shell as the speech noise.
We evaluate \systemname{} on the two datasets, compared to the latest baselines: DPRNN \cite{luo2020dual}, TFGridNet \cite{wang2023tf}. We report the average performance of the two datasets in the Table. \ref{table5}. 
Besides, we present the SNR improvement, against the input noisy audio (5dB) in Fig. \ref{fig:eval_abcs} and \ref{fig:eval_emsb} for each speaker of the dataset. We observe that our model works well for most of the speakers and obtains around 10 dB improvement, indicating a similar performance on our self-collected dataset.
}

\subsection{Adaptive Speech Enhancement}
\begin{figure}
    \centering
    \begin{minipage}{0.48\linewidth}
        \centering
        \includegraphics[width=1\linewidth]{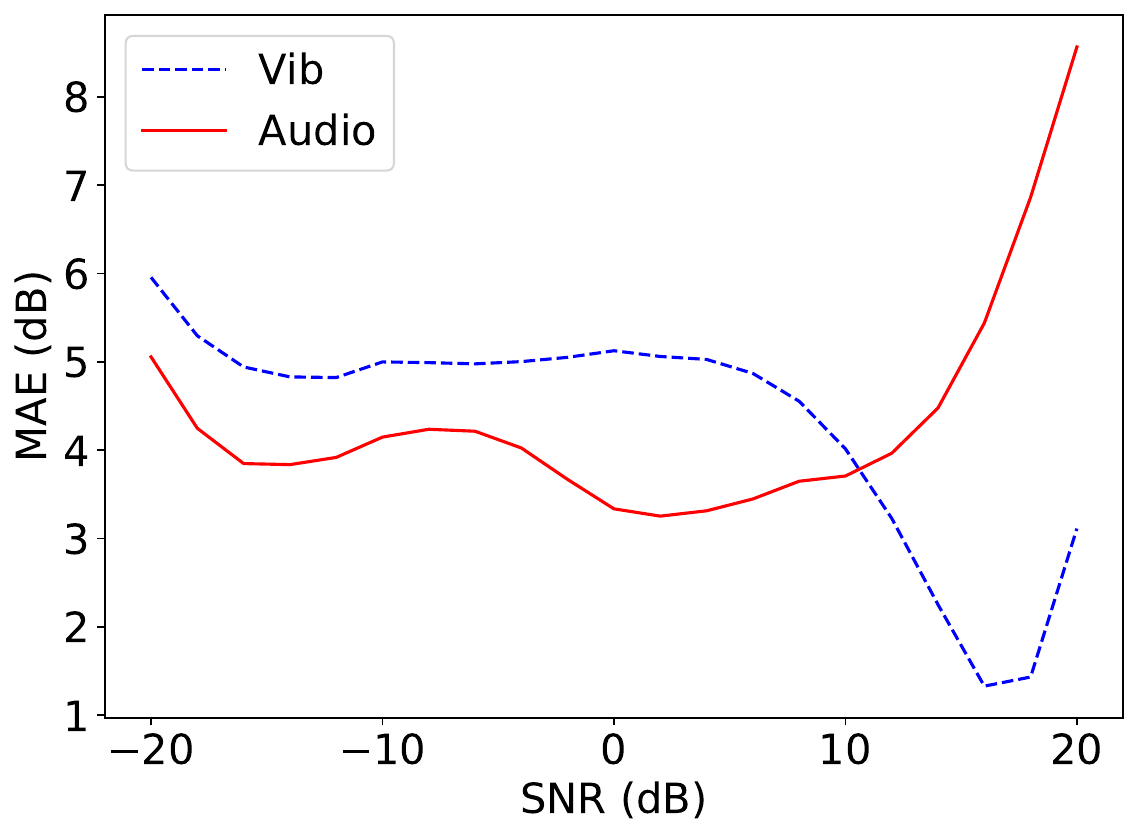}
        \caption{SNR estimation errors.}
        \label{fig:snr_estimation}
    \end{minipage}\hfill
    \begin{minipage}{0.48\linewidth}
        \centering
        \includegraphics[width=1\linewidth]{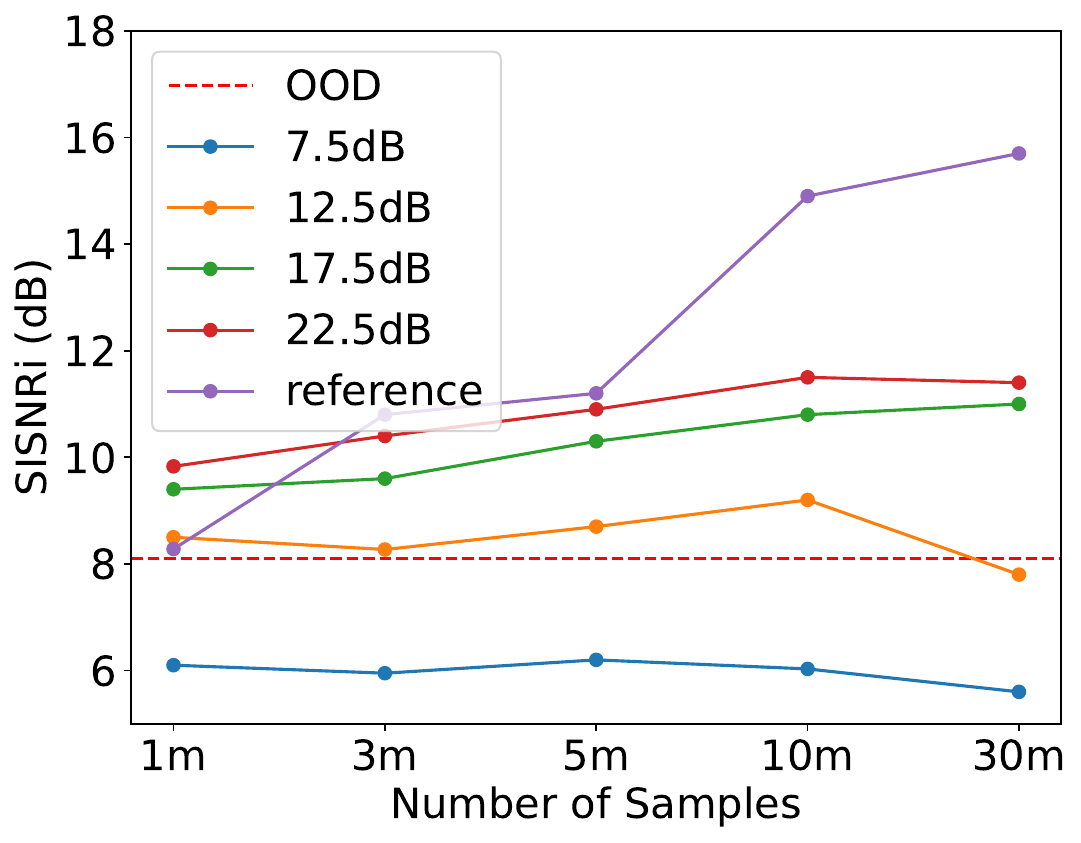}
        \caption{Continual learning performance vs. SNR threshold}
        \label{fig:continual_learning_eval}
    \end{minipage}
    \vspace{-1.5em}
\end{figure}
\new{
To evaluate the continual learning, we evaluate it from two perspectives: 1) the performance of SNR estimation, and 2) the performance of continual learning.

\noindent\textbf{SNR Estimation} is assessed using the mean average error (MAE) between the estimated SNR and the ground truth. It is important to note that the output is constrained to the range of possible SNR values (-20dB to 20dB). In Fig. \ref{fig:snr_estimation}, we compare the proposed SNR estimation method with the audio-only SNR estimation \cite{subakan2021realm}.
Our observations indicate that the benefits of multi-modal learning are not as pronounced when the input SNR is relatively low ($SNR < 0dB$); however, the estimation accuracy significantly improves when the input SNR is high. In contrast, the audio-only SNR estimation struggles in high SNR conditions, as it fails to distinguish between the user's speech and similar background interferences. Since the continual learning is interested in finding the clean audio (high SNR), our proposed SNR estimator is much better than the baseline.

\noindent\textbf{Adaptive training} is evaluated at various SNR thresholds. Specifically, we classify data samples with an estimated SNR above the threshold as "clean" data. It's important to note that while a higher threshold indicates more reliable data, it also results in a lower proportion of effective data. As illustrated in Fig. \ref{fig:continual_learning_eval}, when we set the threshold at 7.5dB or 12.5dB, the training process fails due to the presence of noisy data. In contrast, with thresholds of 17.5dB and 22.5dB, continual learning achieves an approximate 3dB improvement compared to the out-of-domain model, all without requiring any clean data.

\begin{figure}
    \centering
    \begin{minipage}{0.48\linewidth}
        \centering
        \includegraphics[width=1\linewidth]{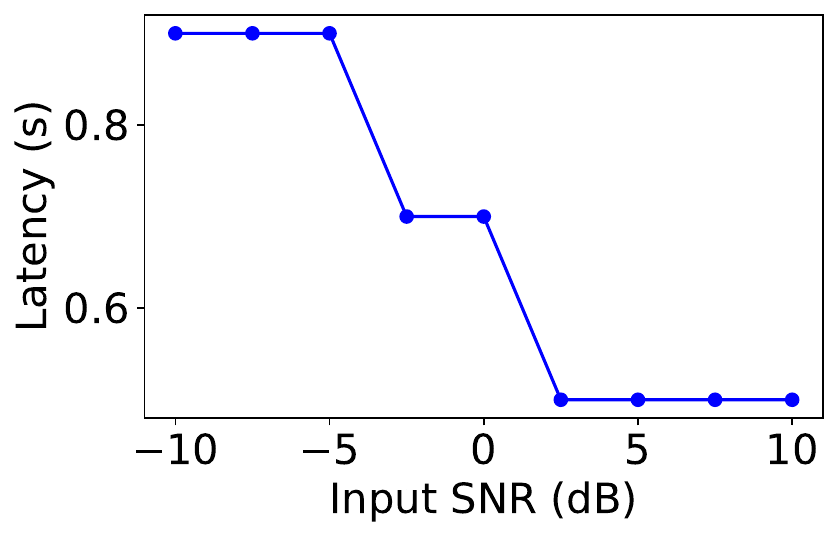}
        \caption{Adaptive speech enhancement latency.}
        \label{fig:adaptive_inference}
    \end{minipage}\hfill
    \begin{minipage}{0.48\linewidth}
        \centering
        \includegraphics[width=1\linewidth]{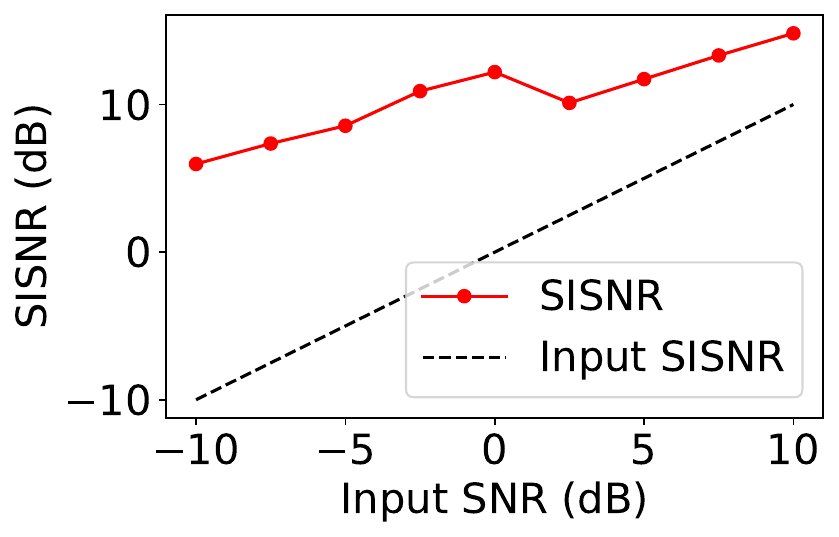}
        \caption{Adaptive speech enhancement performance.}
        \label{fig:adaptive_performance}
    \end{minipage}
    \vspace{-1.5em}
\end{figure}

\noindent\textbf{Adaptive Inference} is evaluated in terms of latency and corresponding effectiveness, as shown in Fig. \ref{fig:adaptive}. As the input SNR increases, \systemname{} dynamically adjusts the model's depth, which helps to reduce computational latency. By default, we set the SNR threshold at 15 dB, meaning that latency begins to decrease once this threshold is surpassed, as indicated in Figs. \ref{fig:adaptive_inference} and \ref{fig:adaptive_performance}.
In Fig. \ref{fig:adaptive_inference}, it is demonstrated that \systemname{} automatically changes the model's depth twice, successfully reducing the latency from 0.9 seconds to 0.5 seconds while maintaining the same output SNR.
}
\begin{figure}
    \centering
    \includegraphics[width=0.75\linewidth]{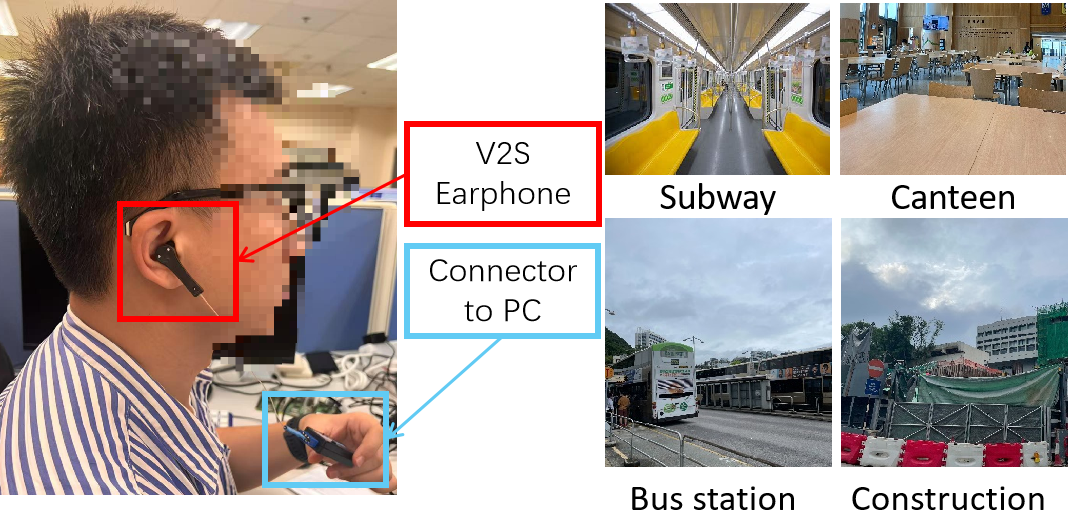}
    \caption{Volunteers with Knowles V2S200D (left) and locations where we collect the dataset (right).}
    \label{fig:new_dataset}
    \vspace{-2em}
\end{figure}

\subsection{In-the-wild Evaluation}
\label{sec:user-study}

\new{
We have developed an advanced prototype of our earables using Knowles' V2S200D Voice Vibration Sensor development kit \cite{V2S200D_Voice_Vibration_Sensor}, which enables dual-channel audio recording (capturing both audio and vibration signals) at a 48 kHz sampling rate. Compared to the prototype described in Sec. \ref{sec:background}, this version offers greater user-friendliness, making it suitable for conducting user studies.

\begin{figure}
    \centering
    \begin{minipage}{0.48\linewidth}
        \centering
        \includegraphics[width=1\linewidth]{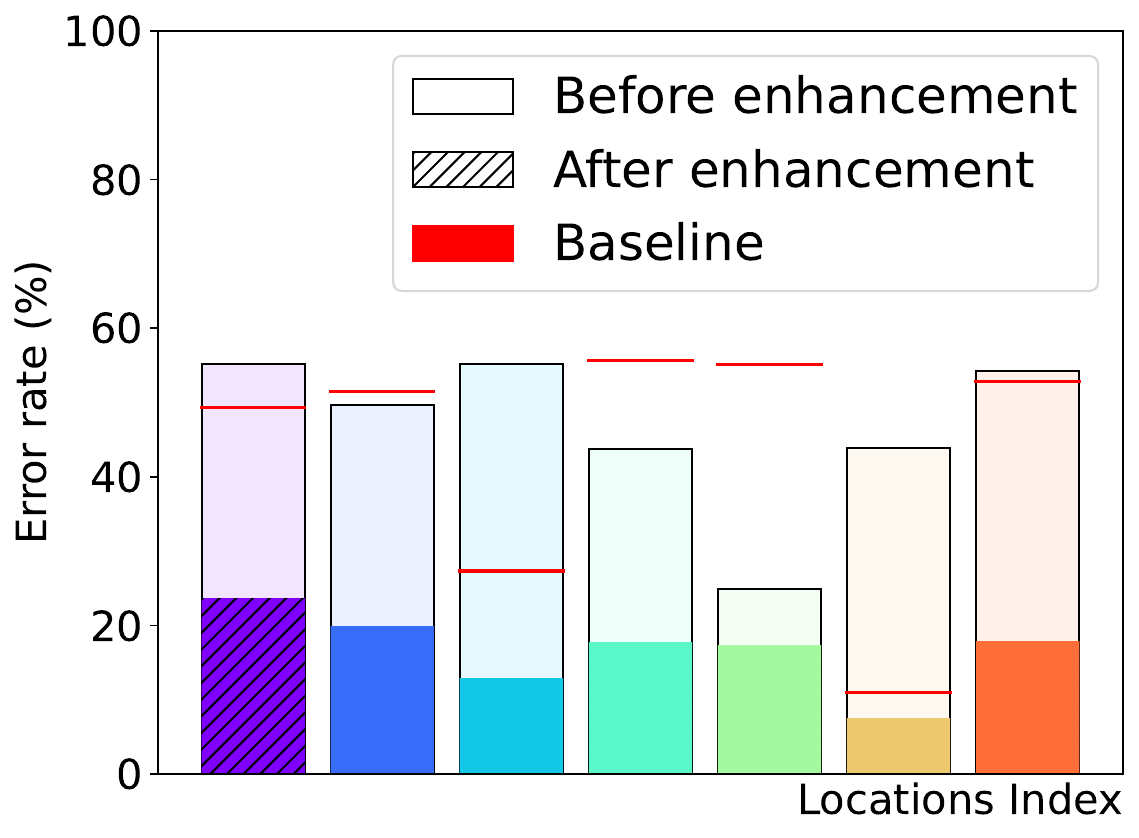}
        \caption{In-the-wild evaluation per location.}
        \label{fig:wild_evaluation_locatiom}
    \end{minipage}\hfill
    \begin{minipage}{0.48\linewidth}
        \centering
        \includegraphics[width=1\linewidth]{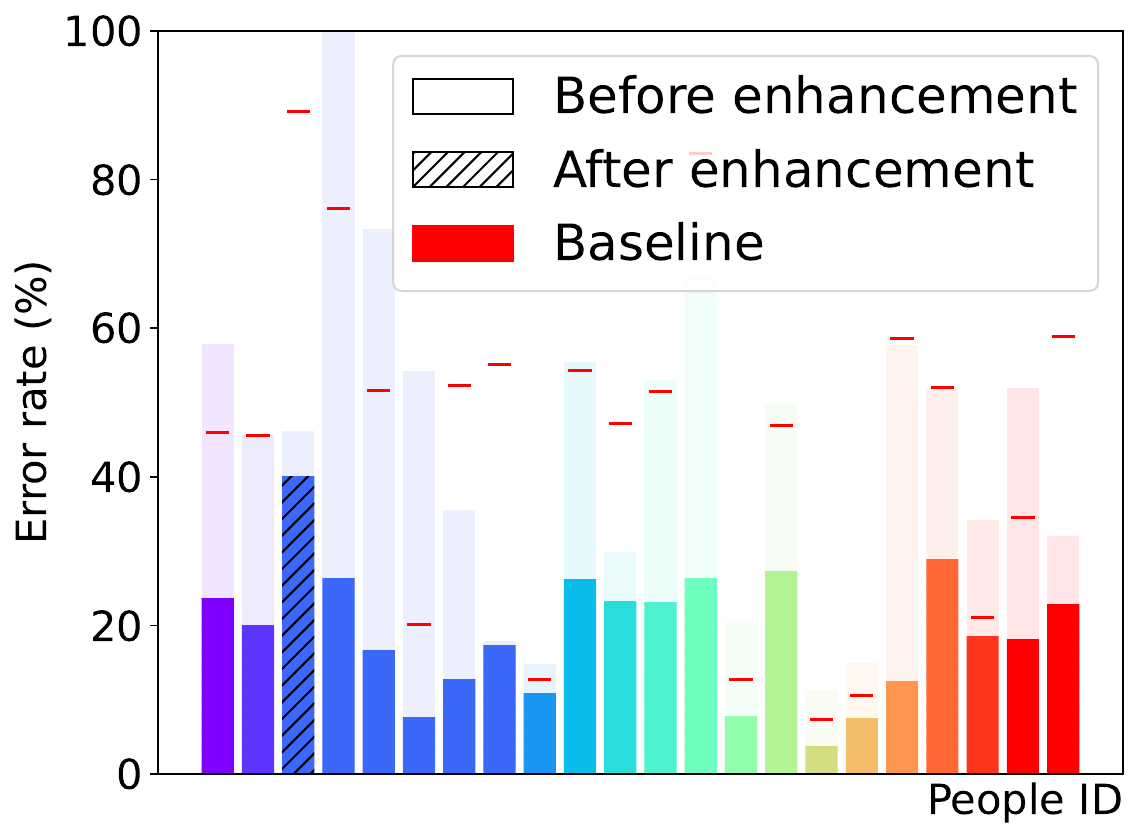}
        \caption{In-the-wild evaluation per speaker.}
        \label{fig:wild_evaluation_speaker}
    \end{minipage}
    \vspace{-1.5em}
\end{figure}

\noindent\textbf{Dataset.}
Specifically, we conduct extensive experiments to validate the performance of \systemname{} in the presence of ongoing noise in real-world environments. Volunteers are asked to read the same content as in the experiments described in Sec. \ref{sec:overall_eval}. In total, our dataset includes 22 speakers. We collect 10 minutes of data from each volunteer at each location. The data is gathered from seven different locations, where each speaker may appear in multiple settings. Specifically, the locations include: restaurant, roadside, coffee bar, office, park, subway, and bus.

\noindent\textbf{Metric.} 
Unlike synthetic noisy speech, we can neither capture the ground truth of clean speech nor evaluate the metrics like SNR and PESQ, nor train the model. Instead, we use the result of Automatic Speech Recognition \footnote{https://github.com/speechbrain/speechbrain} to evaluate the quality of the speech. We use Word Error Rate ($\text{WER}=\frac{S+D+I}{N}$) as the evaluation metric, where S is the number of substitutions, D is the number of deletions, I is the number of insertions, and N is the number of words in the reference. The higher the value of WER, the lower the audio quality. Considering the dataset is in Mandarin, we calculate the Character-Error-Rate in the same way as WER.

\noindent\textbf{Results.} 
As shown in Fig. \ref{fig:wild_evaluation_speaker} and Fig. \ref{fig:wild_evaluation_locatiom}, \systemname{} outperforms both the baseline and the input noisy audio by a large margin in most cases. Specifically, only in one place \systemname{} performs similarly to the baseline, which is the roadside, where the noise is less significant. And we do observe that the improvement from baseline is not obvious for 3 people because the audio is not so noisy, so it is not necessary to have good noise removal. In total, \systemname{} gains 44\% less word error rate compared to the baseline.
}

\subsection{User Study}

\noindent\textbf{Questionnaire Design}.
We recruited 35 volunteers to evaluate the perceived performance of \systemname{}.
Volunteers listened to 5-second audio clips from the dataset in Section 5.1. In the first part, they transcribed audio enhanced by \systemname{} to assess intelligibility using Word Error Rate (WER). In the second part, they compared pairs of audio clips with identical content, selecting the higher-quality audio in two scenarios: \systemname{} versus original noisy audio and \systemname{} versus the baseline (FSN). Each comparison was repeated five times. We measured \systemname{}’s improvement using the correct ratio $\frac{P}{P+N}$, where $P$ is the number of times \systemname{} was preferred and $N$ is the alternative.

\noindent\textbf{Study results}.
According to the results of the first part, \systemname{} achieves an overall WER of $21.5\%$, which is acceptable for understanding the audio content and confirms the effectiveness of \systemname{}.
According to the answers to the second question, the survey results show that $87\%$ of the participants choose \systemname{} over the baseline, and $72\%$ of them choose \systemname{} over the original audio without any enhancements.
In addition, we discuss with participants why they prefer the original audio sounds over the baseline. 
The baseline can produce acoustic artifacts and sometimes wrongly suppress the sound of the target speaker. We note that some participants observed that the impact of artifacts and suppression is hindered after knowing the content or listening repeatedly. However, the speech generated by the baseline causes lots of misunderstanding for the first-time listener. In conclusion, the user study results show that \systemname{} can enhance speech quality and improve user experience compared with the original audio and the baseline.

\section{Discussion}
% Towards deployment \systemname{} on the commercial device, we have the following considerations:
\new{
\begin{itemize}
    \item 
    The current wireless communication in earables does not support two-channel audio recording by default. \systemname{} requires a bit rate of 153.6 kbps (6 × 16 bits × 1.6 kHz) to transmit acceleration data to a mobile device without compression, which is below the maximal bandwidth for Bluetooth 5.0. However, the compatibility with existing profiles and IMU data compression is necessary.
    \item 
    To save the computation resource, we can offload neural network inference on a smartphone, but this still brings additional energy consumption from the ADC (0.54 mW). In comparison, each AirPods Pro earbud has a 43 mAh battery, where VibVoice adds only approximately 1.5\% to the power consumption of earphones. 
    \item 
    The quality of vibration data in earables depends on the hardware form factor and sensor placement. Based on extensive evaluations, we recommend two guidelines for optimal IMU sensor placement: 1) position sensors close to the vibrating organ, and 2) ensure tight contact with the head. Additionally, considerations for user comfort and compatibility with existing devices are crucial.    
    \item 
    We envision integrating \systemname{} with on-device processing to enhance compatibility (eliminating the need for smartphone software installation) and reduce overhead. Devices like OmniBuds \footnote{https://www.omnibuds.tech/} already support on-device machine learning, making this a promising avenue to explore.
\end{itemize}
}

% \section*{Acknowledgments}
% This should be a simple paragraph before the References to thank those individuals and institutions who have supported your work on this article.

\bibliographystyle{IEEEtran}
\bibliography{reference}

% \newpage
 \vspace{-3em}

\begin{IEEEbiography}[{\includegraphics[width=1in,height=1.25in,clip,keepaspectratio]{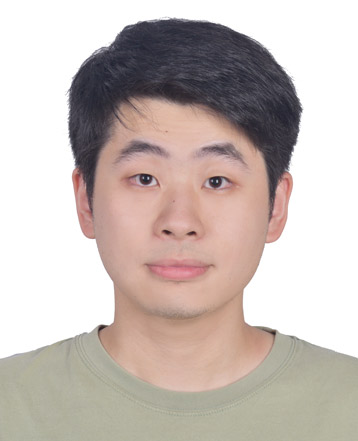}}]{Lixing He} received the B.S. degree in automation from UESTC, Chengdu, China, in 2021, and is a Ph.D. student at Embedded AI and IoT Lab (AIoT Lab), Department of Information Engineering, The Chinese University of Hong Kong. His research interests include audio, wearables technology, and human-centric sensing,

\end{IEEEbiography}

 \vspace{-3.5em}

\begin{IEEEbiography}[{\includegraphics[width=1in,height=1.25in,clip,keepaspectratio]{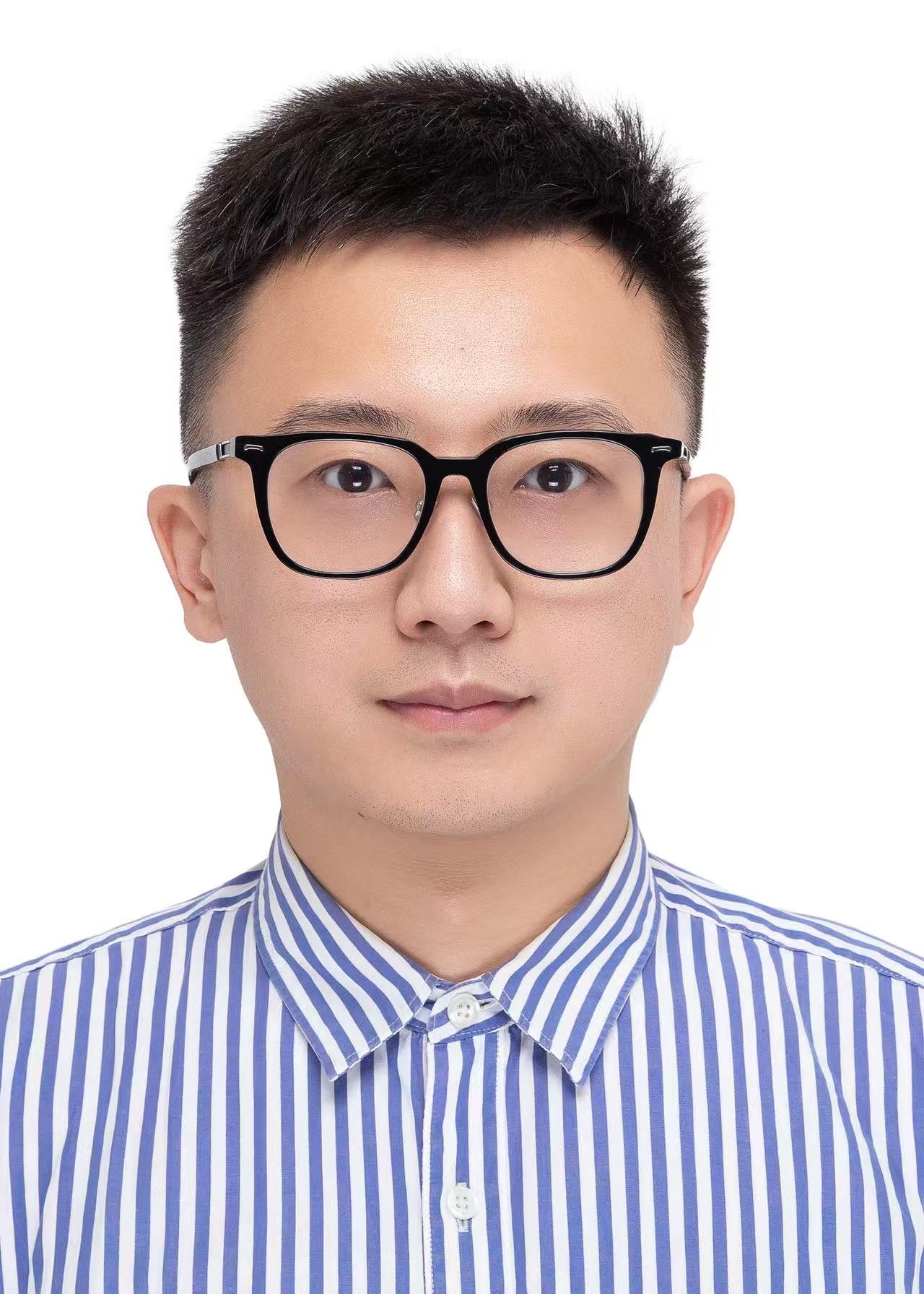}}]{Yunqi Guo} received the B.S. degree in computer science from Shanghai Jiao Tong University, Shanghai, China, in 2016, and the M.S. and Ph.D. degrees in computer science from the University of California, Los Angeles, Los Angeles, CA, USA, in 2018 and 2023, advised by Prof. Songwu Lu. He is currently a Postdoctoral Fellow at The Chinese University of Hong Kong, working with Prof. Guoliang Xing. His research interests lie at the intersection of augmented reality, mobile systems, visual–language interaction, and intelligent sensing. 
\end{IEEEbiography}

 \vspace{-3.5em}

\begin{IEEEbiography}[{\includegraphics[width=1in,height=1.25in,clip,keepaspectratio]{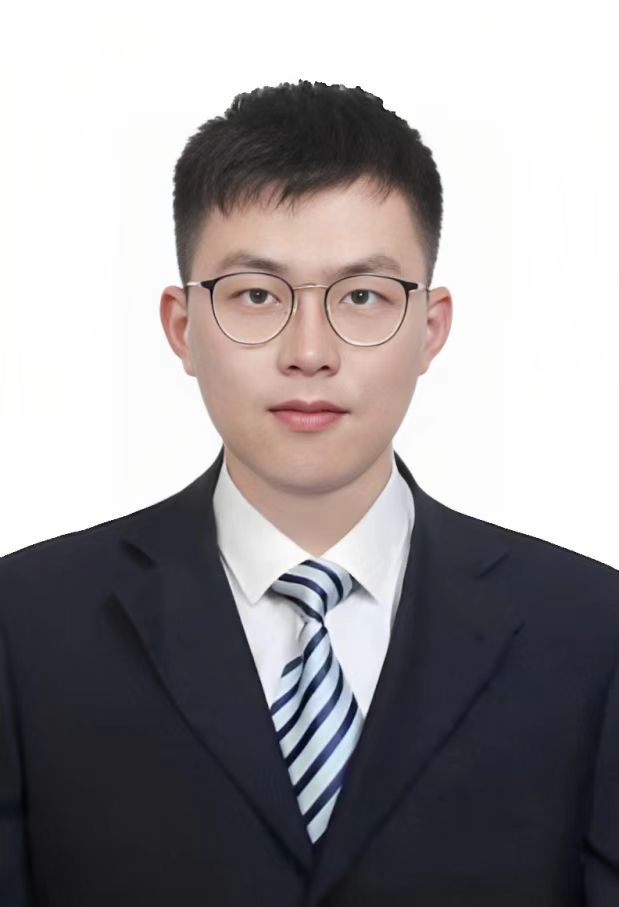}}]{Haozheng Hou} is a Ph.D. student at Embedded AI and IoT Lab (AIoT Lab), Department of Information Engineering, The Chinese University of Hong Kong. He received his B.S. degree (2021) from Nanjing University. His research interests include underwater acoustic sensing and human activity recognition.
\end{IEEEbiography}

 \vspace{-3.5em}

\begin{IEEEbiography}[{\includegraphics[width=1in,height=1.25in,clip,keepaspectratio]{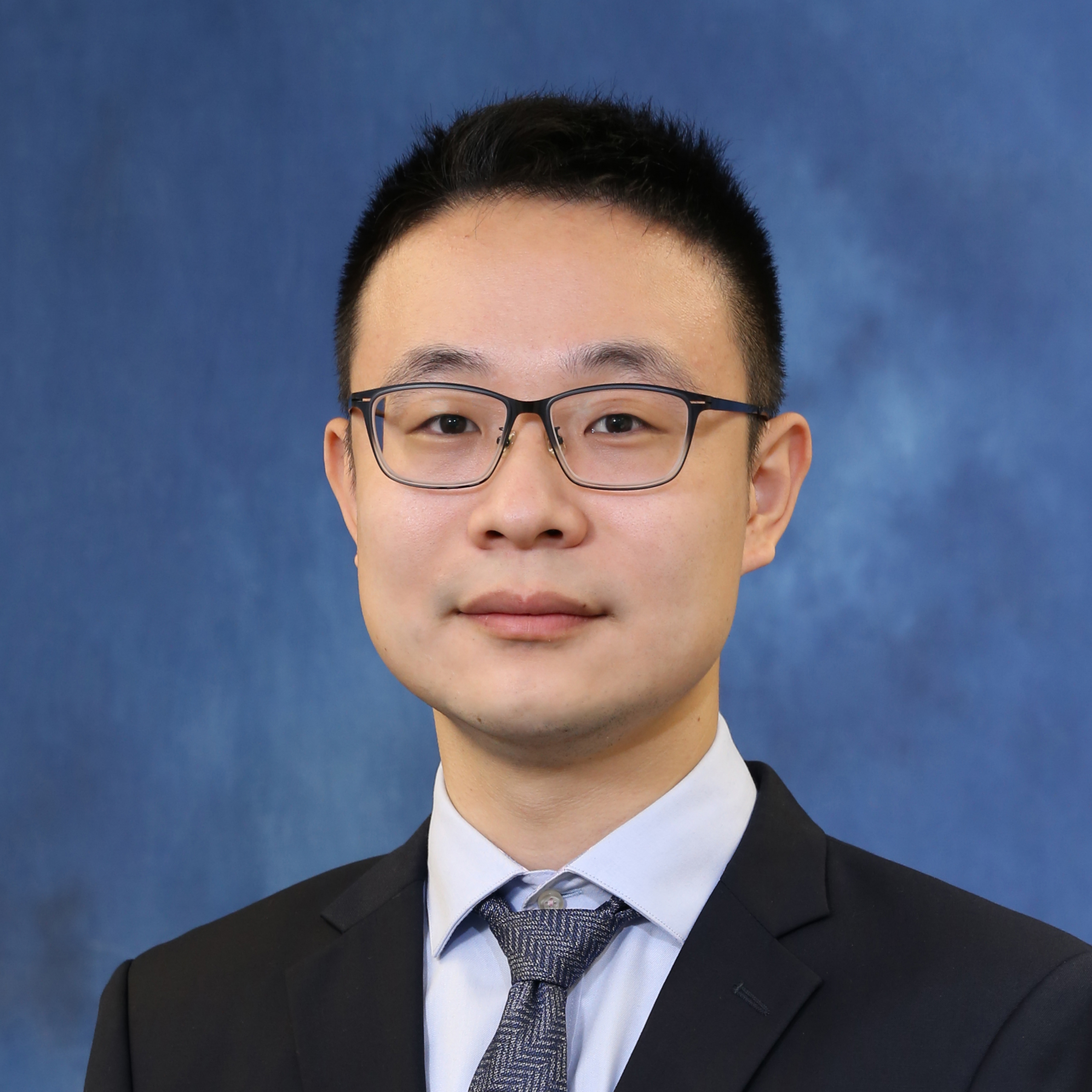}}]{Zhenyu Yan} is an Assistant Professor at The Chinese University of Hong Kong. Dr. Yan has extensive experience in sensing systems, signal and information processing, cyber-physical systems, and machine learning in IoT systems. His works have been published in top international conferences and journals, such as MobiCom, SenSys, IPSN, IEEE Transactions on Mobile Computing, and ACM Transactions on Sensor Networks. He is the recipient of the Rising Star Award from ACM SIGBED China. His papers also received the Best Community Contributions Award at ACM MobiCom 2023, the Best Paper Award Runner-up at ACM MobiCom 2022, and the Best Artifact Award Runner-up at ACM/IEEE IPSN 2021.
\end{IEEEbiography}


\vfill

\end{document}